\documentclass[aps,prl,amsmath,amssymb,twocolumn,superscriptaddress,showpacs]{revtex4-1}
\usepackage{graphicx}				
\usepackage{amssymb}
\usepackage{amsmath}

\usepackage{color}

\newcommand{\Ham}{\hat{H}}
\newcommand{\hc}{\textrm{h.c.}}

\begin{document}
\title{Light-induced d-wave superconductivity\\ through
Floquet-engineered Fermi surfaces in cuprates}

\author{Dante M.~Kennes}
\affiliation{Dahlem Center for Complex Quantum Systems and Fachbereich Physik, Freie Universit\"at Berlin, 14195 Berlin, Germany}
\author{Martin Claassen}
\affiliation{Center for Computational Quantum Physics (CCQ), The Flatiron Institute, 162 Fifth Avenue, New York NY 10010}
\author{Michael A.Sentef}
\affiliation{Max Planck Institute for the Structure and Dynamics of Matter,
Center for Free Electron Laser Science, 22761 Hamburg, Germany}
\author{Christoph Karrasch}
\affiliation{Dahlem Center for Complex Quantum Systems and Fachbereich Physik, Freie Universit\"at Berlin, 14195 Berlin, Germany}

\begin{abstract}
We introduce a mechanism for light-induced Floquet engineering of the Fermi surface to dynamically tip the balance between competing instabilities in correlated condensed matter systems in the vicinity of a van-Hove singularity. We first calculate how the Fermi surface is deformed by an off-resonant, high-frequency light field and then determine the impact of this deformation on the ordering tendencies using an unbiased functional renormalization group approach. As a testbed, we investigate Floquet engineering in cuprates driven by light. We find that the $d$-wave superconducting ordering tendency in this system can be strongly enhanced over the Mott insulating one. This gives rise to extended regions of induced $d$-wave superconductivity in the effective phase diagram in the presence of a light field. 
\end{abstract}

\maketitle

\textit{Introduction---}
The selective manipulation of quantum many-body systems by external driving fields is a blossoming research field bridging nonequilibrium dynamics in cold atoms \cite{Jotzu2014} and condensed matter \cite{Basov2017a,Buzzi2018}. A developing paradigm in ultrafast materials science is \textit{nonequilibrium materials engineering} by means of quasi-periodic pump laser excitation. This opens up a zoo of opportunities to coherently control and switch materials' properties by acting specifically on electronic and lattice degrees of freedom to control band structures \cite{OWNGedik1,OWNGedik2} and their topology \cite{Oka2009,Lindner2011,Sentef2015a,Hubener2017a,Topp2018a}, phonons and lattice deformations \cite{Forst2011,Subedi2014a,VonHoegen2018}, phases with conventional \cite{Kemper2015,Sentef2016b,Knap2016a,Komnik2016b,
Kennes2017b,Babadi2017a,Murakami2017a,Puviani2018} and topological \cite{Grushin2014,Claassen2017a} order, and effective couplings \cite{Singla2015,Pomarico2017,Sentef2017c,Kennes2017b,
Coulthard2017a,Claassen2016,Kitamura2017,Kennes2018c,Kennes2018b,Tancogne-Dejean2017c}. In particular, resonant control of materials' properties on ultrafast time scales in ordered phases has been discussed regarding excitation and manipulation of collective modes \cite{Schmitt2008,Matsunaga2013,Mansart2013,Krauser2014,Murakami2016,Sentef2017b}. While resonant excitation enables selectivity through the choice of driving frequency, it also usually leads to strong heating effects. 
Consequently, resonant ultrafast materials engineering requires accounting for highly non-thermal distributional changes as well as materials details beyond effective low-energy models, both of which are still outstanding challenges for strongly correlated systems.
Conceptually, off-resonant and in particular high-frequency driving offers a simpler pathway to ultrafast materials control via Floquet engineering of effective Hamiltonians \cite{Bukov2015c,Abanin2017}.

Here, we introduce Floquet engineering of the Fermi surface by off-resonant, high-frequency laser driving as a mechanism to enable ultrafast control of the phase diagram in a correlated material. As a showcase example, we investigate a driven version of the two-dimensional three-band Emery model \cite{Emery1987} relevant for cuprate high-temperature superconductors \cite{Keimer2015}. In a first step, we show that off-resonant laser driving with linearly polarized light along crystal axes modifies the Fermi surface of a downfolded one-band Hubbard model, unidirectionally increasing the density of states. Secondly, an unbiased functional renormalization group analysis of the Floquet-engineered two-dimensional Hubbard model is performed. We find as the central result of this paper that off-resonant Fermi surface deformation enables selective tuning between competing antiferromagnetic Mott insulating and d-wave superconducting instabilities through the laser amplitude. Our proposal paves the way to reach d-wave superconductivity even at half-filling, possibly enabling light-induced unconventional superconductivity in undoped parent compounds of cuprate superconductors. 

\textit{Model---}
Whereas the low-energy physics in cuprates is canonically assumed to be well-described by an effective single-band Hubbard model with weak frustration, a proper description of the light-matter interaction with a high-frequency pump must necessarily start from a model including the multi-orbital character of the Cu-O planes. A minimal description of a charge-transfer insulator at half filling is given by the three-band Emery model of Cu $d_{x^2-y^2}$ and O $p_x, p_y$ orbitals \cite{Emery1987}. Including the pump field introduced via Peierls substitution this model reads:
\begin{align}
	H &= -t_{\textrm{pd}} \sum_{n\sigma} \left[ \hat{d}^\dag_{n,\sigma} \left( \hat{p}_{n+x,\sigma} e^{iA_x(t)} - \hat{p}_{n-x,\sigma} e^{-iA_x(t)} \right) \right.   \notag\\
	& \left. ~~~~~+ \hat{d}^\dag_{n,\sigma} \left( \hat{p}_{n+y,\sigma} e^{iA_y(t)} - \hat{p}_{n-y,\sigma} e^{-iA_y(t)} \right) + \textrm{h.c.} \right] \notag\\
	&- t_{\textrm{pp}} \sum_{n\sigma} \left[ \hat{p}^\dag_{n+y,\sigma} \left( \hat{p}_{n-x,\sigma} e^{-i(A_y(t)+A_x(t))} \right. \right. \notag\\
	& \left. \left. ~~~~~~~ - \hat{p}^\dag_{n+x} e^{-i(A_y(t) - A_x(t))} \right) + \textrm{h.c.} \right] \notag\\
	&+ U_{\textrm{dd}} \sum_n \hat{n}^d_{n\uparrow} \hat{n}^d_{n\downarrow} + U_{\textrm{pp}} \sum_m \hat{n}^p_{m\uparrow} \hat{n}^p_{m\downarrow} + \Delta \sum_{m\sigma} \hat{n}^p_{m\sigma}   \label{eq:DrivenEmeryModel}
\end{align}
Here, $\hat{d}$, $\hat{p}$ denote annihilation operators for electrons on the Cu $d_{x^2-y^2}$ orbital and O $p_x, p_y$ orbitals, respectively, and $n,m$ sum over Cu and O sites respectively. The pump field $\mathbf{A}(t) = \left[ A_x(t),~ A_y(t) \right]^\top$ enters via minimal substitution for the hopping matrix elements, which are parameterized by $t_{pd}$ for hopping between Cu and O sites, and $t_{pp}$ for hybridization between $p_x$, $p_y$ orbitals on neighboring O sites. In the following, we consider a linearly-polarized pump pulse $\mathbf{A}(t) = [ A^{\rm Light} \cos(\Omega t), ~0 ]^\top$. Here, $A^{\rm Light} = a_0 e \mathcal{E} / (\hbar \Omega)$ is the dimensionless pump strength, with $a_0$, $e$ and $\mathcal{E}$ the lattice constant, electron charge and electric field, respectively. Furthermore, $U_{dd}$, $U_{pp}$ denote local Coulomb repulsion for Cu and O, and $\Delta$ denotes the charge transfer energy.

If the pump frequency $\Omega$ is larger than the bandwidth of Eq. (\ref{eq:DrivenEmeryModel}) and chosen to be off-resonant from higher-energy interband transitions, heating can be neglected on short (pre-thermalized) time scales \cite{Bukov2015c,Mentink2015a,Claassen2017a,Kennes2018b,Peronaci2018}. Instead, for $\Omega \to \infty$, short-time effective dynamics can be described via a time-averaged Hamiltonian $H_{\textrm{eff}} = (1/T) \int_0^T H(t)$, which corresponds to the zeroth order contribution of a high-frequency expansion of the Floquet Hamiltonian \cite{Magnus1954a}. In this case, in analogy to dynamical bond softening in molecular optics \cite{Bucksbaum1990}, the dominant effect of the light field is to selectively renormalize effective hoppings via choice of pump polarization \cite{Eckardt2017a}. Crucially, for a linearly-polarized pump pulse, this entails an \textit{anisotropic} photo-induced renormalization of the hopping matrix elements, dynamically breaking the lattice rotation symmetries. In the high-frequency limit, the three-band Emery model thus acquires anisotropic Cu-O hoppings $t_{\rm pd}^x = t^{\rm eq}_{\rm pd} \mathcal{J}_0(A)$, $t_{\rm pd}^y = t^{\rm eq}_{\rm pd}$, as well as a renormalization of the O-O hopping $t_{\rm pp} = t_{\rm pp}^{\rm eq} \mathcal{J}_0(A)$, where $t^{\rm eq}_{\rm pd}$, $t^{\rm eq}_{\rm pp}$ denote the equilibrium values.
%
To achieve the downfolding to a single band Hubbard model we proceed from the period-averaged anisotropic Emery model and adopt a two-step strategy of first integrating out the O degrees of freedom and deriving an effective spin model of Cu local moments Eq. (\ref{eq:DrivenEmeryModel}) and then proposing an equivalent anisotropic single-band Hubbard model that reproduces these Cu-Cu spin-exchange interactions. 

After diagonalizing the O-O hybridization $t_{pp}$ and expanding to fourth order in $t_{pd}$, the effective spin exchange interactions read [see supplementary information]
\begin{widetext}
\begin{align}
	J^y &=  \frac{4 (t^y_{\text{pd}})^2 \left(\Delta  t^y_{\text{pd}}-t^x_{\text{pd}} t_{\text{pp}}\right){}^2}{U_{\text{dd}} \left(t_{\text{pp}}^2-\Delta ^2\right){}^2} + \frac{4 \Delta  t^x_{\text{pd}} (t^y_{\text{pd}})^3 t_{\text{pp}} \left(6 \Delta +U_{\text{pp}}\right)}{\left(t_{\text{pp}}-\Delta \right){}^2 \left(\Delta +t_{\text{pp}}\right){}^2 \left(2 t_{\text{pp}}^2-\Delta  \left(2 \Delta +U_{\text{pp}}\right)\right)} \notag\\
	&+ \frac{8 t^x_{\text{pd}} (t^y_{\text{pd}})^3 t_{\text{pp}}^3}{\left(t_{\text{pp}}-\Delta \right){}^2 \left(\Delta +t_{\text{pp}}\right){}^2 \left(2 t_{\text{pp}}^2-\Delta  \left(2 \Delta +U_{\text{pp}}\right)\right)} - \frac{16 \Delta  (t^x_{\text{pd}})^2 (t^y_{\text{pd}})^2 t_{\text{pp}}^2}{\left(t_{\text{pp}}-\Delta \right){}^2 \left(\Delta +t_{\text{pp}}\right){}^2 \left(2 t_{\text{pp}}^2-\Delta  \left(2 \Delta +U_{\text{pp}}\right)\right)} \notag\\
	&+ \frac{2 (t^y_{\text{pd}})^4 t_{\text{pp}}^2 \left(8 \Delta ^2+8 \Delta  U_{\text{pp}}+U_{\text{pp}}^2\right)}{\left(t_{\text{pp}}^2-\Delta ^2\right){}^2 \left(2 \Delta +U_{\text{pp}}\right) \left(\Delta  \left(2 \Delta +U_{\text{pp}}\right)-2 t_{\text{pp}}^2\right)}+\frac{8 \Delta ^3 (t^y_{\text{pd}})^4}{\left(t_{\text{pp}}^2-\Delta ^2\right){}^2 \left(\Delta  \left(2 \Delta +U_{\text{pp}}\right)-2 t_{\text{pp}}^2\right)} \\
	J^x &= \left. J^y \right|_{ t^x_{\text{pd}} \leftrightarrow t^y_{\text{pd}}  } \\
	J' &= \frac{4 (t^x_{\text{pd}})^2 (t^y_{\text{pd}})^2 t_{\text{pp}}^2 \left(2 t_{\text{pp}}^2-\Delta  \left(2 \Delta +4 U_{\text{dd}}+U_{\text{pp}}\right)\right)}{U_{\text{dd}} \left(t_{\text{pp}}^2-\Delta ^2\right){}^2 \left(2 t_{\text{pp}}^2-\Delta  \left(2 \Delta +U_{\text{pp}}\right)\right)}
\end{align}
\end{widetext}
where $J^{x,y}$ denote anisotropic nearest-neighbor spin-exchange on the Cu square lattice, and $t_{pp}$ furthermore stabilizes next-nearest-neighbor spin exchange $J'$.

\begin{figure*}[t]
\centering
\includegraphics[width=\columnwidth]{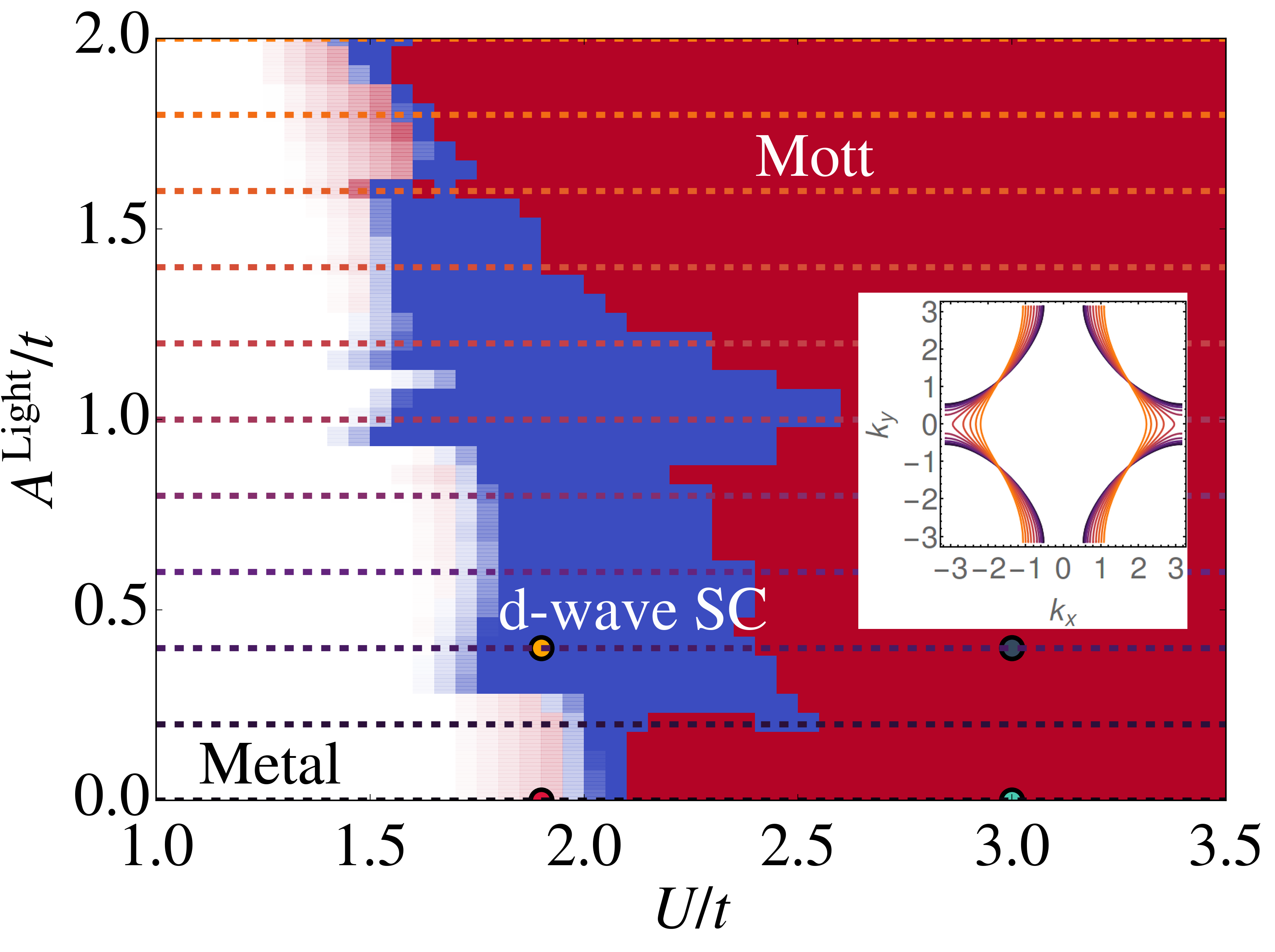}
\includegraphics[width=\columnwidth]{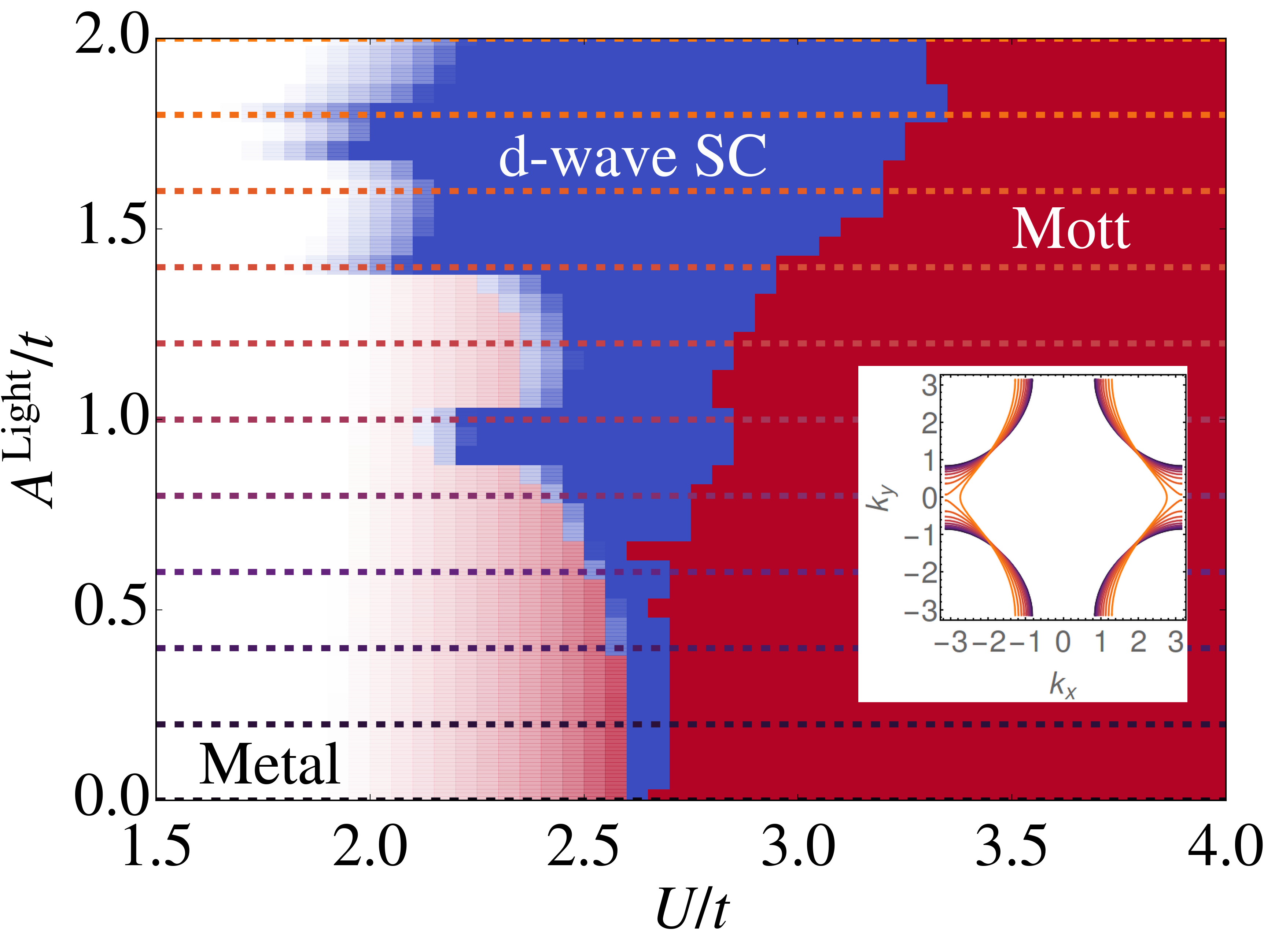}
\caption{\textit{Main Panels:} Phase diagrams of light induced phases in the cuprates in dependence of the strength of the light field $A^{\rm Light}$ as well the Hubbard interaction $U$. Blue and red indicate whether d-wave superconducting or Mott insulting tendencies are dominant, respectively. The more saturated the color, the stronger the pairing instability is in either phase. \textit{Left:} Results at half filling $n=1$. Colored dots indicated the paramters used in the different panels of Fig.~\ref{fig:Res_vertex}. \textit{Right:} Results for electron-doped systems  $n\approx1.2$. \textit{Insets:} Deformation of the Fermi surfaces with increasing strength of the light field (the corresponding values of $A^{\rm Light}$ are indicated by horizontal dashed lines in the main panel). }
\label{fig:Res_PD}
\end{figure*}

We choose canonical parameters $t^{\rm eq}_{\rm pd} = 1.13eV$, $t^{\rm eq}_{\rm pp} = 0.49eV$, $\Delta = 3.24eV$, $U_{\rm dd} = 8.5eV$ and $U_{\rm pp} = 4.1eV$ for the Cu-O plane \cite{Ohta1991,Chen2010c,Chen2013}. For La$_2$CuO$_4$ as a prototypical undoped cuprate parent compound, thus a dimensionless field strength of $A^{\rm Light} = 1$ amounts to an electric field strength of $\mathcal{E} = 125 ~\textrm{mV/\AA}$ for a $3 eV$ pump laser. 

\begin{figure}[t]
\centering
\includegraphics[width=0.9\columnwidth]{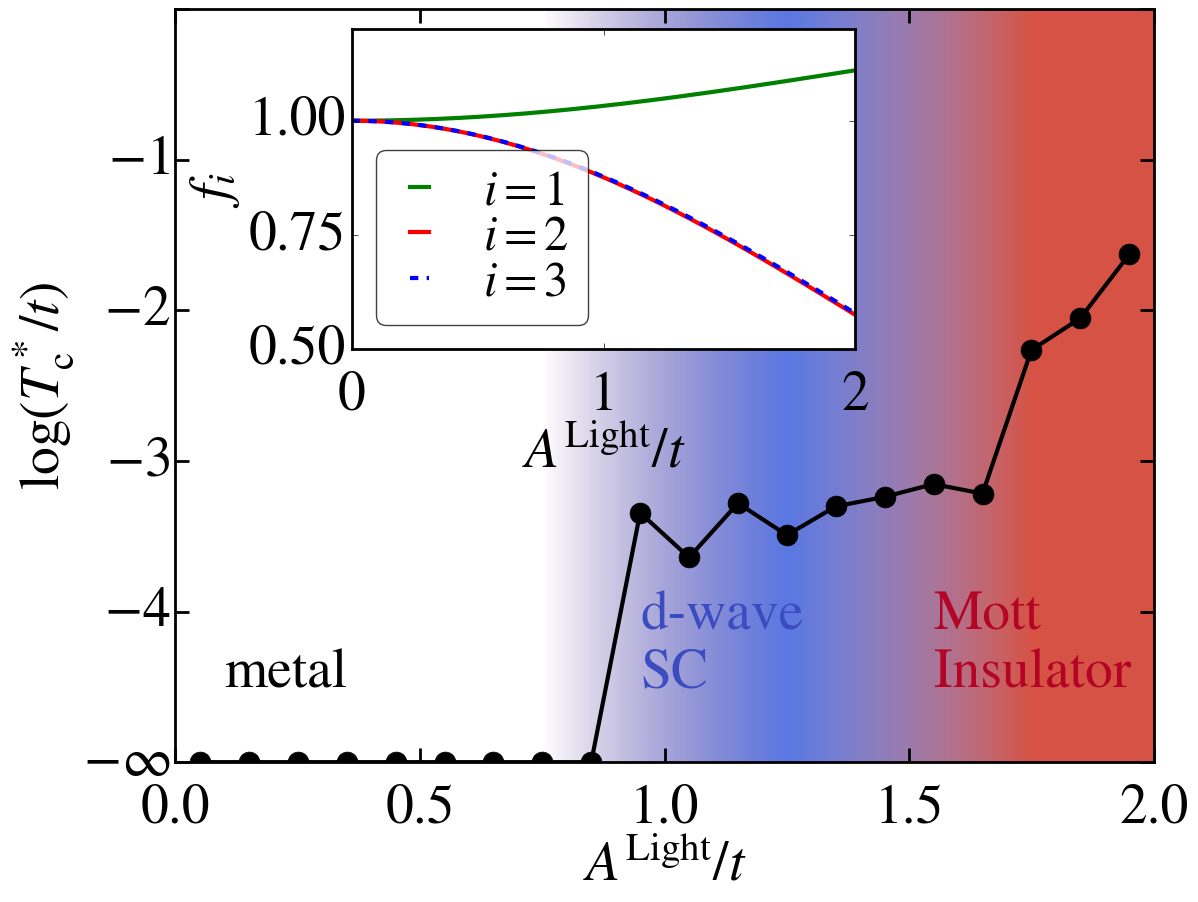}
\caption{\textit{Main Panel:} Critical temperature $T_\textnormal{c}^*$ for the different d-wave superconducting and Mott ordering tendencies at half-filling ($n=1$) and $U/t=1.75$ as a function of the light field strength $A^{\rm Light}$. As $A^{\rm Light}$ is increased, the d-wave superconducting phase is first stabilized over the metallic one, and at even larger $A^{\rm Light}$ a transition into a Mott insulating state occurs. \textit{Inset:} The three functions $f_1(A^{\rm Light})$, $f_2(A^{\rm Light})$, and $f_3(A^{\rm Light})$ describing the field-dependent changes in the Hamiltonian of the cuprates, Eq.~\eqref{eq:H}.}
\label{fig:Res_TcN05}
\end{figure}

Crucially, the photo-induced anisotropy $t_x / t_y = \sqrt{J^x / J^y} $ and frustration $t' / t_x = \sqrt{J' / J^x}$ of the single-band hopping matrix elements can be extracted without knowledge of the energy scale of the effective single-band Hubbard interaction. The latter does not have a straightforward microscopic derivation and is instead introduced via comparison to experiment. The non-interacting part  of the effective single-band Hamiltonian describing the light-driven cuprates can thus be characterized by the dispersion relation
\begin{align}
	\epsilon(k_x,k_y) = &-2 f_1(A^{\rm Light}) \cos(k_x) - 2 f_2(A^{\rm Light}) \cos(k_y)\notag\\
&+4 t' f_3(A^{\rm Light})\cos(k_x)\cos(k_y)\label{eq:H}
\end{align}
with $t'=0.24831t$ and the three field strength dependent functions $f_1$, $f_2$ and $f_3$ shown in the inset of Fig.~\ref{fig:Res_TcN05}. To describe the effects of correlations we add a Hubbard, onsite density-density interaction with strength $U$.

\textit{Method---}
We employ the functional renormalization group to study the effects of strong correlations induced by the Hubbard interaction $U$. 
Within this method, the many-body problem is recast into an infinite hierarchy of flow equations for the coupling constants with a flow parameter $\Lambda$ which is introduced as an infrared cutoff. We use the implementation described in Ref.~\onlinecite{Metzner2012a}. In a nutshell, the main approximation is the truncation of the aforementioned hierarchy by concentrating solely on the renormalization of the effective two-particle scattering $\gamma_2^\Lambda(\vec k_1,\vec k_2, \vec k_1',\vec k_2')$ with $\gamma_2^{\Lambda_\textnormal{initial}}\sim U$. After discretizing the Brillouin zone into patches, the flow equations can be solved numerically \cite{Metzner2012a,Kennes2018d}.

If during the flow the effective two-body scattering exhibits a divergence at a given $\Lambda_\textnormal{C}$, this is interpeted as an ordering tendency whose precise nature can be identified from the momentum structure of $\gamma_2^{\Lambda_\textnormal{C}}$ (see, e.g., Ref.~\cite{Metzner2012a} as well as the discussion below).

\textit{Results---}
Our main results are summarized in the effective phase diagrams shown in Fig.~\ref{fig:Res_PD} as a function of pump field strength $A^{\rm Light}$ and the Hubbard interaction $U$ for two values of the filling $n$ (half filling $n=1$ and $n=1.2$ on the left and right, respectively). The saturation of the color indicates how strong the ordering tendency is in the following sense: We define $\gamma_2^{\Lambda_\textnormal{C}}/t>25$ as the criterion of a divergence at which we stop the flow and extract the type of order from the symmetry of $\gamma_2^{\Lambda_\textnormal{C}}$ \cite{Metzner2012a}. If the flow can be continued down to $\Lambda=0$ (i.e., in the absence of a divergence) and if also $\gamma_2^{\Lambda=0}/t<5$, we interpret this as a metal. We color the corresponding parts of the phase diagram in a saturated white, blue, and red tone corresponding to metallic, superconducting, and Mott insulating phases, respectively. If the flow shows no divergence but if $5<\gamma_2^{\Lambda=0}/t<25$, we mark this using an unsaturated color which we choose according to the leading ordering tendency associated with $\gamma_2^{\Lambda=0}$. This protocol accounts for the ambiguity in the definition of the term `divergence'.

In equilibrium ($A^{\rm Light}=0$), the metallic state for a frustrated Fermi surface is unstable towards forming a Mott insulator  beyond a critical interaction strength, with d-wave superconductivity preempting the metal-insulator transition for a very narrow parameter regime. As the light field is turned on ($A^{\rm Light}>0$), the superconducting regions are enlarged significantly, which one can understand as follows: A finite $A^{\rm Light}>0$ deforms the effective Floquet Fermi surface (see insets to Fig.~\ref{fig:Res_PD}), which is successively tuned closer to the van-Hove points $(k_x, k_y)=(\pm \pi,0)$ until at a critical field strength $A_{\rm van-Hove}^{\rm Light}$, the van-Hove singularity falls onto the Fermi surface (we find $A_{\rm van-Hove}^{\rm Light} \approx t$ and $A_{\rm van-Hove}^{\rm Light} \approx 1.8 t$ for $n=1$ and $n=1.2$, respectively). As the van-Hove points are approached by the effective Fermi surface, the fragile balance between Mott insulating behavior (which is triggered by approximate nesting) and d-wave superconductivity (which is triggered by the vicinity to the van-Hove points) is tipped towards the latter. Correspondingly, optical pumping considerably broadens the d-wave superconducting regime that intervenes between the metal and the Mott insulator. Therefore, we predict that by increasing the light field $A^{\rm Light}$ one can tune either from a metal or a Mott insulating state into a d-wave superconducting phase, traversing a vertical line in the effective phase diagrams of Fig.~\ref{fig:Res_PD}.

\begin{figure}[t]
\centering
\includegraphics[width=0.9\columnwidth]{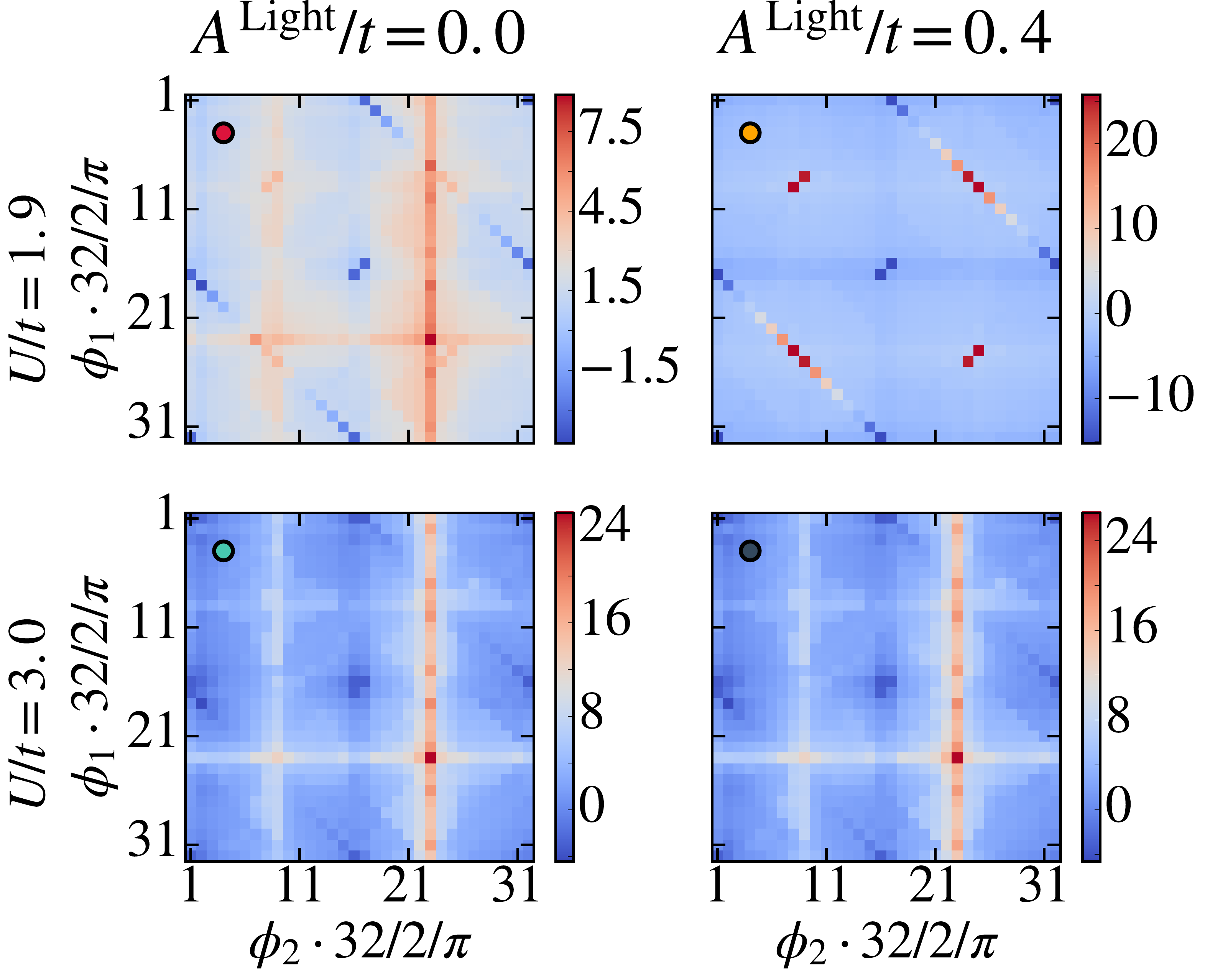}
\caption{Typical results for the effective interaction vertex $\gamma_2^{\Lambda_\textnormal{C}}$ in dependence of the angles $\phi_{1,2}$ of two incoming momenta $\vec k_{1,2}$ on the Fermi surface. The angle of the third (outgoing) momentum is held fixed at $\phi_3=2\pi/32$. The results are obtained for half-filling ($n=1$), but the doped case looks qualitatively similar. For ease of orientation, a colored dot in the upper left corner indicates the corresponding parameters in the left panel of Fig.~\ref{fig:Res_PD}. The lack of a divergence (upper left panel) is reflective of a metallic phase. A divergence with dominant vertical features (lower panels) indicate Mott ordering tendencies, while dominant diagonal features with alternating signs (upper right panel) are associated with d-wave superconducting order (see \cite{Metzner2012a}). For $U=1.9$ (upper panels), we find that turning on a light field can induce a d-wave superconductivity ordering tendency.  }
\label{fig:Res_vertex}
\end{figure}

Explicit results for the two-particle scattering at the end of the flow are presented in Fig.~\ref{fig:Res_vertex} for the four combinations of $A^{\rm Light}/t=0,0.4$ with $U/t=1.9,3$. We plot $\gamma_2$ as a function of the angle of two incoming momenta for a fixed third outgoing momentum, all of which lie on the Fermi surface. Deep in the Mott phase ($U/t=3$), the light field does not alter the structure of the two-particle vertex. However, for intermediate interaction strength ($U/t=1.9$), turning on the light field strongly promotes d-wave superconducting behavior, which is reflected by a pronounced diagonal structure in the vertex of alternating positive and negative values \cite{Metzner2012a}. 

Finally, we identify the cutoff energy scale $\Lambda_\textnormal{c}$ at which the flow diverges with a critical temperature $T_\textnormal{c}^*=\Lambda_\textnormal{C}$ below which strong ordering tendencies are expected. This identification, though rough, was shown to yield results which are qualitatively consistent with those obtained by setting up an RG scheme at finite physical temperature \cite{Metzner2012a}. In Fig.~\ref{fig:Res_PD}, we plot the critical temperatures as a function of $A^{\rm Light}$ for a fixed interaction strength $U/t=1.75$ at half filling, which permits tuning from a metal to d-wave superconductivity to a Mott insulator. We find a clear plateau in the critical temperature at about $T_\textnormal{c}/t=0.001$ when the systems enters into the d-wave superconducting state, with a further jump in $T_\textnormal{c}$ upon transitioning into the Mott insulating state at even higher $A^{\rm Light}\approx1.6$.

\textit{Discussion and Outlook---}
We report on light-induced d-wave superconducting behavior in driven cuprates. The mechanism we propose relies on the Floquet modification of the hoppings and as a consequence of the Fermi surface, which in turn can tune the fragile balance between competing orders present in these compounds. We find that the domains of light-induced d-wave superconductivity extend considerably into the parameter regimes which in equilibrium are Mott insulators. This is related to shifting the Fermi surface closer to the van-Hove points (which tends to stabilize d-wave superconducting behavior) and to weakening approximate nesting (which favors the competing Mott insulating ordering tendency). Both of these mechanism thus conspire to extend the d-wave superconducting phase in Fig.~\ref{fig:Res_PD} at finite light-field strength. 

To ascertain this behavior, we have studied quasi-equilibrium steady states of photo-modulated Hubbard models in a high-frequency expansion. Two crucial open questions for future work entail the role of heating to determine time scales beyond which energy absorption dwarfs the reported ordering tendencies as well as the detailed dynamical mechanism of crossing a Mott-superconductor phase transition out of equilibrium. The former likely necessitates taking into account lattice and higher-energy band effects, whereas the latter opens up intriguing avenues of investigation regarding Kibble-Zurek scaling and defects generated as a function of the speed by which the Fermi surface is tuned by the light field \cite{Kennes2018c}. The competition between heating and defect generation rates likely determines an optimal pump width. In either case, methods to describe the driven non-equilibrium physics of strongly correlated two-dimensional systems are still scarce, and a proper treatment of the heating effects is currently beyond theoretical reach.

Despite these open issues the effect is remarkably robust and can be seen at different fillings as well as for different systems. As the mechanism relies only on the fact that the underlying system exhibits competing instabilities, which is true in a variety of correlated materials, it should be equally applicable to other interesting materials systems, such as the ruthenates \cite{Maeno1994} or twisted bilayer graphene \cite{Cao2018c}.

\textit{Acknowledgements---}
D.M.K. and C.K. acknowledge support by the Deutsche Forschungsgemeinschaft through the Emmy Noether program (KA 3360/2-1). M.C. acknowledges support from the Flatiron Institute, a division of the Simons Foundation. M.A.S. acknowledges financial support by the DFG through the Emmy Noether program (SE 2558/2-1). Simulations were performed with computing resources granted by RWTH Aachen University under projects rwth0013 and prep0010.

\bibliographystyle{apsrev4-1}
\bibliography{LIDSC}

\begin{thebibliography}{54}%
\makeatletter
\providecommand \@ifxundefined [1]{%
 \@ifx{#1\undefined}
}%
\providecommand \@ifnum [1]{%
 \ifnum #1\expandafter \@firstoftwo
 \else \expandafter \@secondoftwo
 \fi
}%
\providecommand \@ifx [1]{%
 \ifx #1\expandafter \@firstoftwo
 \else \expandafter \@secondoftwo
 \fi
}%
\providecommand \natexlab [1]{#1}%
\providecommand \enquote  [1]{``#1''}%
\providecommand \bibnamefont  [1]{#1}%
\providecommand \bibfnamefont [1]{#1}%
\providecommand \citenamefont [1]{#1}%
\providecommand \href@noop [0]{\@secondoftwo}%
\providecommand \href [0]{\begingroup \@sanitize@url \@href}%
\providecommand \@href[1]{\@@startlink{#1}\@@href}%
\providecommand \@@href[1]{\endgroup#1\@@endlink}%
\providecommand \@sanitize@url [0]{\catcode `\\12\catcode `\$12\catcode
  `\&12\catcode `\#12\catcode `\^12\catcode `\_12\catcode `\%12\relax}%
\providecommand \@@startlink[1]{}%
\providecommand \@@endlink[0]{}%
\providecommand \url  [0]{\begingroup\@sanitize@url \@url }%
\providecommand \@url [1]{\endgroup\@href {#1}{\urlprefix }}%
\providecommand \urlprefix  [0]{URL }%
\providecommand \Eprint [0]{\href }%
\providecommand \doibase [0]{http://dx.doi.org/}%
\providecommand \selectlanguage [0]{\@gobble}%
\providecommand \bibinfo  [0]{\@secondoftwo}%
\providecommand \bibfield  [0]{\@secondoftwo}%
\providecommand \translation [1]{[#1]}%
\providecommand \BibitemOpen [0]{}%
\providecommand \bibitemStop [0]{}%
\providecommand \bibitemNoStop [0]{.\EOS\space}%
\providecommand \EOS [0]{\spacefactor3000\relax}%
\providecommand \BibitemShut  [1]{\csname bibitem#1\endcsname}%
\let\auto@bib@innerbib\@empty
\bibitem [{\citenamefont {Jotzu}\ \emph {et~al.}(2014)\citenamefont {Jotzu},
  \citenamefont {Messer}, \citenamefont {Desbuquois}, \citenamefont {Lebrat},
  \citenamefont {Uehlinger}, \citenamefont {Greif},\ and\ \citenamefont
  {Esslinger}}]{Jotzu2014}%
  \BibitemOpen
  \bibfield  {author} {\bibinfo {author} {\bibfnamefont {G.}~\bibnamefont
  {Jotzu}}, \bibinfo {author} {\bibfnamefont {M.}~\bibnamefont {Messer}},
  \bibinfo {author} {\bibfnamefont {R.}~\bibnamefont {Desbuquois}}, \bibinfo
  {author} {\bibfnamefont {M.}~\bibnamefont {Lebrat}}, \bibinfo {author}
  {\bibfnamefont {T.}~\bibnamefont {Uehlinger}}, \bibinfo {author}
  {\bibfnamefont {D.}~\bibnamefont {Greif}}, \ and\ \bibinfo {author}
  {\bibfnamefont {T.}~\bibnamefont {Esslinger}},\ }\href {\doibase
  10.1038/nature13915} {\bibfield  {journal} {\bibinfo  {journal} {Nature}\
  }\textbf {\bibinfo {volume} {515}},\ \bibinfo {pages} {237} (\bibinfo {year}
  {2014})},\ \Eprint {http://arxiv.org/abs/1406.7874} {arXiv:1406.7874}
  \BibitemShut {NoStop}%
\bibitem [{\citenamefont {Basov}\ \emph {et~al.}(2017)\citenamefont {Basov},
  \citenamefont {Averitt},\ and\ \citenamefont {Hsieh}}]{Basov2017a}%
  \BibitemOpen
  \bibfield  {author} {\bibinfo {author} {\bibfnamefont {D.~N.}\ \bibnamefont
  {Basov}}, \bibinfo {author} {\bibfnamefont {R.~D.}\ \bibnamefont {Averitt}},
  \ and\ \bibinfo {author} {\bibfnamefont {D.}~\bibnamefont {Hsieh}},\ }\href
  {\doibase 10.1038/nmat5017} {\bibfield  {journal} {\bibinfo  {journal}
  {Nature Materials}\ }\textbf {\bibinfo {volume} {16}},\ \bibinfo {pages}
  {1077} (\bibinfo {year} {2017})}\BibitemShut {NoStop}%
\bibitem [{\citenamefont {Buzzi}\ \emph {et~al.}(2018)\citenamefont {Buzzi},
  \citenamefont {F{\"{o}}rst}, \citenamefont {Mankowsky},\ and\ \citenamefont
  {Cavalleri}}]{Buzzi2018}%
  \BibitemOpen
  \bibfield  {author} {\bibinfo {author} {\bibfnamefont {M.}~\bibnamefont
  {Buzzi}}, \bibinfo {author} {\bibfnamefont {M.}~\bibnamefont {F{\"{o}}rst}},
  \bibinfo {author} {\bibfnamefont {R.}~\bibnamefont {Mankowsky}}, \ and\
  \bibinfo {author} {\bibfnamefont {A.}~\bibnamefont {Cavalleri}},\ }\href
  {\doibase 10.1038/s41578-018-0024-9} {\ ,\ \bibinfo {pages} {1} (\bibinfo
  {year} {2018})},\ \Eprint {http://arxiv.org/abs/1805.02376}
  {arXiv:1805.02376} \BibitemShut {NoStop}%
\bibitem [{\citenamefont {Mahmood}\ \emph {et~al.}(2016)\citenamefont
  {Mahmood}, \citenamefont {Chan}, \citenamefont {Alpichshev}, \citenamefont
  {Gardner}, \citenamefont {Lee}, \citenamefont {Lee},\ and\ \citenamefont
  {Gedik}}]{OWNGedik1}%
  \BibitemOpen
  \bibfield  {author} {\bibinfo {author} {\bibfnamefont {F.}~\bibnamefont
  {Mahmood}}, \bibinfo {author} {\bibfnamefont {C.-k.}\ \bibnamefont {Chan}},
  \bibinfo {author} {\bibfnamefont {Z.}~\bibnamefont {Alpichshev}}, \bibinfo
  {author} {\bibfnamefont {D.}~\bibnamefont {Gardner}}, \bibinfo {author}
  {\bibfnamefont {Y.}~\bibnamefont {Lee}}, \bibinfo {author} {\bibfnamefont
  {P.~A.}\ \bibnamefont {Lee}}, \ and\ \bibinfo {author} {\bibfnamefont
  {N.}~\bibnamefont {Gedik}},\ }\href {\doibase 10.1038/NPHYS3609} {\bibfield
  {journal} {\bibinfo  {journal} {Nature Physics}\ }\textbf {\bibinfo {volume}
  {12}},\ \bibinfo {pages} {306} (\bibinfo {year} {2016})}\BibitemShut
  {NoStop}%
\bibitem [{\citenamefont {Wang}\ \emph {et~al.}(2014)\citenamefont {Wang},
  \citenamefont {Steinberg}, \citenamefont {Jarillo-Herrero},\ and\
  \citenamefont {Gedik}}]{OWNGedik2}%
  \BibitemOpen
  \bibfield  {author} {\bibinfo {author} {\bibfnamefont {Y.}~\bibnamefont
  {Wang}}, \bibinfo {author} {\bibfnamefont {H.}~\bibnamefont {Steinberg}},
  \bibinfo {author} {\bibfnamefont {P.}~\bibnamefont {Jarillo-Herrero}}, \ and\
  \bibinfo {author} {\bibfnamefont {N.}~\bibnamefont {Gedik}},\ }\href
  {\doibase 10.1364/UP.2014.10.Thu.A.2} {\bibfield  {journal} {\bibinfo
  {journal} {19th International Conference on Ultrafast Phenomena}\ }\textbf
  {\bibinfo {volume} {342}},\ \bibinfo {pages} {10.Thu.A.2} (\bibinfo {year}
  {2014})},\ \Eprint {http://arxiv.org/abs/1310.7563} {arXiv:1310.7563}
  \BibitemShut {NoStop}%
\bibitem [{\citenamefont {Oka}\ and\ \citenamefont {Aoki}(2009)}]{Oka2009}%
  \BibitemOpen
  \bibfield  {author} {\bibinfo {author} {\bibfnamefont {T.}~\bibnamefont
  {Oka}}\ and\ \bibinfo {author} {\bibfnamefont {H.}~\bibnamefont {Aoki}},\
  }\href {\doibase 10.1103/PhysRevB.79.081406} {\bibfield  {journal} {\bibinfo
  {journal} {Physical Review B - Condensed Matter and Materials Physics}\
  }\textbf {\bibinfo {volume} {79}},\ \bibinfo {pages} {081406} (\bibinfo
  {year} {2009})},\ \Eprint {http://arxiv.org/abs/0807.4767} {arXiv:0807.4767}
  \BibitemShut {NoStop}%
\bibitem [{\citenamefont {Lindner}\ \emph {et~al.}(2011)\citenamefont
  {Lindner}, \citenamefont {Refael},\ and\ \citenamefont
  {Galitski}}]{Lindner2011}%
  \BibitemOpen
  \bibfield  {author} {\bibinfo {author} {\bibfnamefont {N.~H.}\ \bibnamefont
  {Lindner}}, \bibinfo {author} {\bibfnamefont {G.}~\bibnamefont {Refael}}, \
  and\ \bibinfo {author} {\bibfnamefont {V.}~\bibnamefont {Galitski}},\ }\href
  {\doibase 10.1038/nphys1926} {\bibfield  {journal} {\bibinfo  {journal}
  {Nature Physics}\ }\textbf {\bibinfo {volume} {7}},\ \bibinfo {pages} {490}
  (\bibinfo {year} {2011})},\ \Eprint {http://arxiv.org/abs/1008.1792}
  {arXiv:1008.1792} \BibitemShut {NoStop}%
\bibitem [{\citenamefont {Sentef}\ \emph {et~al.}(2015)\citenamefont {Sentef},
  \citenamefont {Claassen}, \citenamefont {Kemper}, \citenamefont {Moritz},
  \citenamefont {Oka}, \citenamefont {Freericks},\ and\ \citenamefont
  {Devereaux}}]{Sentef2015a}%
  \BibitemOpen
  \bibfield  {author} {\bibinfo {author} {\bibfnamefont {M.~A.}\ \bibnamefont
  {Sentef}}, \bibinfo {author} {\bibfnamefont {M.}~\bibnamefont {Claassen}},
  \bibinfo {author} {\bibfnamefont {A.~F.}\ \bibnamefont {Kemper}}, \bibinfo
  {author} {\bibfnamefont {B.}~\bibnamefont {Moritz}}, \bibinfo {author}
  {\bibfnamefont {T.}~\bibnamefont {Oka}}, \bibinfo {author} {\bibfnamefont
  {J.~K.}\ \bibnamefont {Freericks}}, \ and\ \bibinfo {author} {\bibfnamefont
  {T.~P.}\ \bibnamefont {Devereaux}},\ }\href {\doibase 10.1038/ncomms8047}
  {\bibfield  {journal} {\bibinfo  {journal} {Nature Communications}\ }\textbf
  {\bibinfo {volume} {6}},\ \bibinfo {pages} {7047} (\bibinfo {year} {2015})},\
  \Eprint {http://arxiv.org/abs/1401.5103} {arXiv:1401.5103} \BibitemShut
  {NoStop}%
\bibitem [{\citenamefont {H{\"{u}}bener}\ \emph {et~al.}(2017)\citenamefont
  {H{\"{u}}bener}, \citenamefont {Sentef}, \citenamefont {{De Giovannini}},
  \citenamefont {Kemper},\ and\ \citenamefont {Rubio}}]{Hubener2017a}%
  \BibitemOpen
  \bibfield  {author} {\bibinfo {author} {\bibfnamefont {H.}~\bibnamefont
  {H{\"{u}}bener}}, \bibinfo {author} {\bibfnamefont {M.~A.}\ \bibnamefont
  {Sentef}}, \bibinfo {author} {\bibfnamefont {U.}~\bibnamefont {{De
  Giovannini}}}, \bibinfo {author} {\bibfnamefont {A.~F.}\ \bibnamefont
  {Kemper}}, \ and\ \bibinfo {author} {\bibfnamefont {A.}~\bibnamefont
  {Rubio}},\ }\href {\doibase 10.1038/ncomms13940} {\bibfield  {journal}
  {\bibinfo  {journal} {Nature Communications}\ }\textbf {\bibinfo {volume}
  {8}},\ \bibinfo {pages} {13940} (\bibinfo {year} {2017})},\ \Eprint
  {http://arxiv.org/abs/1604.03399} {arXiv:1604.03399} \BibitemShut {NoStop}%
\bibitem [{\citenamefont {Topp}\ \emph {et~al.}(2018)\citenamefont {Topp},
  \citenamefont {Tancogne-Dejean}, \citenamefont {Kemper}, \citenamefont
  {Rubio},\ and\ \citenamefont {Sentef}}]{Topp2018a}%
  \BibitemOpen
  \bibfield  {author} {\bibinfo {author} {\bibfnamefont {G.~E.}\ \bibnamefont
  {Topp}}, \bibinfo {author} {\bibfnamefont {N.}~\bibnamefont
  {Tancogne-Dejean}}, \bibinfo {author} {\bibfnamefont {A.~F.}\ \bibnamefont
  {Kemper}}, \bibinfo {author} {\bibfnamefont {A.}~\bibnamefont {Rubio}}, \
  and\ \bibinfo {author} {\bibfnamefont {M.~A.}\ \bibnamefont {Sentef}},\
  }\href {\doibase arXiv:1803.07447v1} {\  (\bibinfo {year} {2018}),\
  arXiv:1803.07447v1},\ \Eprint {http://arxiv.org/abs/1803.07447}
  {arXiv:1803.07447} \BibitemShut {NoStop}%
\bibitem [{\citenamefont {F{\"{o}}rst}\ \emph {et~al.}(2011)\citenamefont
  {F{\"{o}}rst}, \citenamefont {Manzoni}, \citenamefont {Kaiser}, \citenamefont
  {Tomioka}, \citenamefont {Tokura}, \citenamefont {Merlin},\ and\
  \citenamefont {Cavalleri}}]{Forst2011}%
  \BibitemOpen
  \bibfield  {author} {\bibinfo {author} {\bibfnamefont {M.}~\bibnamefont
  {F{\"{o}}rst}}, \bibinfo {author} {\bibfnamefont {C.}~\bibnamefont
  {Manzoni}}, \bibinfo {author} {\bibfnamefont {S.}~\bibnamefont {Kaiser}},
  \bibinfo {author} {\bibfnamefont {Y.}~\bibnamefont {Tomioka}}, \bibinfo
  {author} {\bibfnamefont {Y.}~\bibnamefont {Tokura}}, \bibinfo {author}
  {\bibfnamefont {R.}~\bibnamefont {Merlin}}, \ and\ \bibinfo {author}
  {\bibfnamefont {A.}~\bibnamefont {Cavalleri}},\ }\href {\doibase
  10.1038/nphys2055} {\bibfield  {journal} {\bibinfo  {journal} {Nature
  Physics}\ }\textbf {\bibinfo {volume} {7}},\ \bibinfo {pages} {854} (\bibinfo
  {year} {2011})},\ \Eprint {http://arxiv.org/abs/1101.1878} {arXiv:1101.1878}
  \BibitemShut {NoStop}%
\bibitem [{\citenamefont {Subedi}\ \emph {et~al.}(2014)\citenamefont {Subedi},
  \citenamefont {Cavalleri},\ and\ \citenamefont {Georges}}]{Subedi2014a}%
  \BibitemOpen
  \bibfield  {author} {\bibinfo {author} {\bibfnamefont {A.}~\bibnamefont
  {Subedi}}, \bibinfo {author} {\bibfnamefont {A.}~\bibnamefont {Cavalleri}}, \
  and\ \bibinfo {author} {\bibfnamefont {A.}~\bibnamefont {Georges}},\ }\href
  {\doibase 10.1103/PhysRevB.89.220301} {\bibfield  {journal} {\bibinfo
  {journal} {Physical Review B - Condensed Matter and Materials Physics}\
  }\textbf {\bibinfo {volume} {89}},\ \bibinfo {pages} {220301} (\bibinfo
  {year} {2014})},\ \Eprint {http://arxiv.org/abs/1311.0544} {arXiv:1311.0544}
  \BibitemShut {NoStop}%
\bibitem [{\citenamefont {{Von Hoegen}}\ \emph {et~al.}(2018)\citenamefont
  {{Von Hoegen}}, \citenamefont {Mankowsky}, \citenamefont {Fechner},
  \citenamefont {F{\"{o}}rst},\ and\ \citenamefont
  {Cavalleri}}]{VonHoegen2018}%
  \BibitemOpen
  \bibfield  {author} {\bibinfo {author} {\bibfnamefont {A.}~\bibnamefont {{Von
  Hoegen}}}, \bibinfo {author} {\bibfnamefont {R.}~\bibnamefont {Mankowsky}},
  \bibinfo {author} {\bibfnamefont {M.}~\bibnamefont {Fechner}}, \bibinfo
  {author} {\bibfnamefont {M.}~\bibnamefont {F{\"{o}}rst}}, \ and\ \bibinfo
  {author} {\bibfnamefont {A.}~\bibnamefont {Cavalleri}},\ }\href {\doibase
  10.1038/nature25484} {\bibfield  {journal} {\bibinfo  {journal} {Nature}\
  }\textbf {\bibinfo {volume} {555}},\ \bibinfo {pages} {79} (\bibinfo {year}
  {2018})},\ \Eprint {http://arxiv.org/abs/1708.07659} {arXiv:1708.07659}
  \BibitemShut {NoStop}%
\bibitem [{\citenamefont {Kemper}\ \emph {et~al.}(2015)\citenamefont {Kemper},
  \citenamefont {Sentef}, \citenamefont {Moritz}, \citenamefont {Freericks},\
  and\ \citenamefont {Devereaux}}]{Kemper2015}%
  \BibitemOpen
  \bibfield  {author} {\bibinfo {author} {\bibfnamefont {A.~F.}\ \bibnamefont
  {Kemper}}, \bibinfo {author} {\bibfnamefont {M.~A.}\ \bibnamefont {Sentef}},
  \bibinfo {author} {\bibfnamefont {B.}~\bibnamefont {Moritz}}, \bibinfo
  {author} {\bibfnamefont {J.~K.}\ \bibnamefont {Freericks}}, \ and\ \bibinfo
  {author} {\bibfnamefont {T.~P.}\ \bibnamefont {Devereaux}},\ }\href {\doibase
  10.1103/PhysRevB.92.224517} {\bibfield  {journal} {\bibinfo  {journal}
  {Physical Review B - Condensed Matter and Materials Physics}\ }\textbf
  {\bibinfo {volume} {92}},\ \bibinfo {pages} {224517} (\bibinfo {year}
  {2015})},\ \Eprint {http://arxiv.org/abs/1412.2762} {arXiv:1412.2762}
  \BibitemShut {NoStop}%
\bibitem [{\citenamefont {Sentef}\ \emph {et~al.}(2016)\citenamefont {Sentef},
  \citenamefont {Kemper}, \citenamefont {Georges},\ and\ \citenamefont
  {Kollath}}]{Sentef2016b}%
  \BibitemOpen
  \bibfield  {author} {\bibinfo {author} {\bibfnamefont {M.~A.}\ \bibnamefont
  {Sentef}}, \bibinfo {author} {\bibfnamefont {A.~F.}\ \bibnamefont {Kemper}},
  \bibinfo {author} {\bibfnamefont {A.}~\bibnamefont {Georges}}, \ and\
  \bibinfo {author} {\bibfnamefont {C.}~\bibnamefont {Kollath}},\ }\href
  {\doibase 10.1103/PhysRevB.93.144506} {\bibfield  {journal} {\bibinfo
  {journal} {Physical Review B}\ }\textbf {\bibinfo {volume} {93}},\ \bibinfo
  {pages} {144506} (\bibinfo {year} {2016})},\ \Eprint
  {http://arxiv.org/abs/1505.07575} {arXiv:1505.07575} \BibitemShut {NoStop}%
\bibitem [{\citenamefont {Knap}\ \emph {et~al.}(2016)\citenamefont {Knap},
  \citenamefont {Babadi}, \citenamefont {Refael}, \citenamefont {Martin},\ and\
  \citenamefont {Demler}}]{Knap2016a}%
  \BibitemOpen
  \bibfield  {author} {\bibinfo {author} {\bibfnamefont {M.}~\bibnamefont
  {Knap}}, \bibinfo {author} {\bibfnamefont {M.}~\bibnamefont {Babadi}},
  \bibinfo {author} {\bibfnamefont {G.}~\bibnamefont {Refael}}, \bibinfo
  {author} {\bibfnamefont {I.}~\bibnamefont {Martin}}, \ and\ \bibinfo {author}
  {\bibfnamefont {E.}~\bibnamefont {Demler}},\ }\href {\doibase
  10.1103/PhysRevB.94.214504} {\bibfield  {journal} {\bibinfo  {journal}
  {Physical Review B}\ }\textbf {\bibinfo {volume} {94}},\ \bibinfo {pages}
  {214504} (\bibinfo {year} {2016})},\ \Eprint
  {http://arxiv.org/abs/1511.07874} {arXiv:1511.07874} \BibitemShut {NoStop}%
\bibitem [{\citenamefont {Komnik}\ and\ \citenamefont
  {Thorwart}(2016)}]{Komnik2016b}%
  \BibitemOpen
  \bibfield  {author} {\bibinfo {author} {\bibfnamefont {A.}~\bibnamefont
  {Komnik}}\ and\ \bibinfo {author} {\bibfnamefont {M.}~\bibnamefont
  {Thorwart}},\ }\href {\doibase 10.1140/epjb/e2016-70528-1} {\bibfield
  {journal} {\bibinfo  {journal} {European Physical Journal B}\ }\textbf
  {\bibinfo {volume} {89}},\ \bibinfo {pages} {244} (\bibinfo {year} {2016})},\
  \Eprint {http://arxiv.org/abs/1607.03858} {arXiv:1607.03858} \BibitemShut
  {NoStop}%
\bibitem [{\citenamefont {Kennes}\ \emph {et~al.}(2017)\citenamefont {Kennes},
  \citenamefont {Wilner}, \citenamefont {Reichman},\ and\ \citenamefont
  {Millis}}]{Kennes2017b}%
  \BibitemOpen
  \bibfield  {author} {\bibinfo {author} {\bibfnamefont {D.~M.}\ \bibnamefont
  {Kennes}}, \bibinfo {author} {\bibfnamefont {E.~Y.}\ \bibnamefont {Wilner}},
  \bibinfo {author} {\bibfnamefont {D.~R.}\ \bibnamefont {Reichman}}, \ and\
  \bibinfo {author} {\bibfnamefont {A.~J.}\ \bibnamefont {Millis}},\ }\href
  {\doibase 10.1038/nphys4024} {\bibfield  {journal} {\bibinfo  {journal}
  {Nature Physics}\ }\textbf {\bibinfo {volume} {13}},\ \bibinfo {pages} {479}
  (\bibinfo {year} {2017})},\ \Eprint {http://arxiv.org/abs/1609.03802v1}
  {arXiv:1609.03802v1} \BibitemShut {NoStop}%
\bibitem [{\citenamefont {Babadi}\ \emph {et~al.}(2017)\citenamefont {Babadi},
  \citenamefont {Knap}, \citenamefont {Martin}, \citenamefont {Refael},\ and\
  \citenamefont {Demler}}]{Babadi2017a}%
  \BibitemOpen
  \bibfield  {author} {\bibinfo {author} {\bibfnamefont {M.}~\bibnamefont
  {Babadi}}, \bibinfo {author} {\bibfnamefont {M.}~\bibnamefont {Knap}},
  \bibinfo {author} {\bibfnamefont {I.}~\bibnamefont {Martin}}, \bibinfo
  {author} {\bibfnamefont {G.}~\bibnamefont {Refael}}, \ and\ \bibinfo {author}
  {\bibfnamefont {E.}~\bibnamefont {Demler}},\ }\href {\doibase
  10.1103/PhysRevB.96.014512} {\bibfield  {journal} {\bibinfo  {journal}
  {Physical Review B}\ }\textbf {\bibinfo {volume} {96}},\ \bibinfo {pages}
  {014512} (\bibinfo {year} {2017})},\ \Eprint
  {http://arxiv.org/abs/1702.02531} {arXiv:1702.02531} \BibitemShut {NoStop}%
\bibitem [{\citenamefont {Murakami}\ \emph {et~al.}(2017)\citenamefont
  {Murakami}, \citenamefont {Tsuji}, \citenamefont {Eckstein},\ and\
  \citenamefont {Werner}}]{Murakami2017a}%
  \BibitemOpen
  \bibfield  {author} {\bibinfo {author} {\bibfnamefont {Y.}~\bibnamefont
  {Murakami}}, \bibinfo {author} {\bibfnamefont {N.}~\bibnamefont {Tsuji}},
  \bibinfo {author} {\bibfnamefont {M.}~\bibnamefont {Eckstein}}, \ and\
  \bibinfo {author} {\bibfnamefont {P.}~\bibnamefont {Werner}},\ }\href
  {\doibase 10.1103/PhysRevB.96.045125} {\bibfield  {journal} {\bibinfo
  {journal} {Physical Review B}\ }\textbf {\bibinfo {volume} {96}},\ \bibinfo
  {pages} {045125} (\bibinfo {year} {2017})},\ \Eprint
  {http://arxiv.org/abs/1702.02942} {arXiv:1702.02942} \BibitemShut {NoStop}%
\bibitem [{\citenamefont {Puviani}\ and\ \citenamefont
  {Sentef}(2018)}]{Puviani2018}%
  \BibitemOpen
  \bibfield  {author} {\bibinfo {author} {\bibfnamefont {M.}~\bibnamefont
  {Puviani}}\ and\ \bibinfo {author} {\bibfnamefont {M.~A.}\ \bibnamefont
  {Sentef}},\ }\href {http://arxiv.org/abs/1806.08187} {\  (\bibinfo {year}
  {2018})},\ \Eprint {http://arxiv.org/abs/1806.08187} {arXiv:1806.08187}
  \BibitemShut {NoStop}%
\bibitem [{\citenamefont {Grushin}\ \emph {et~al.}(2014)\citenamefont
  {Grushin}, \citenamefont {G{\'{o}}mez-Le{\'{o}}n},\ and\ \citenamefont
  {Neupert}}]{Grushin2014}%
  \BibitemOpen
  \bibfield  {author} {\bibinfo {author} {\bibfnamefont {A.~G.}\ \bibnamefont
  {Grushin}}, \bibinfo {author} {\bibfnamefont {{\'{A}}.}~\bibnamefont
  {G{\'{o}}mez-Le{\'{o}}n}}, \ and\ \bibinfo {author} {\bibfnamefont
  {T.}~\bibnamefont {Neupert}},\ }\href {\doibase
  10.1103/PhysRevLett.112.156801} {\bibfield  {journal} {\bibinfo  {journal}
  {Physical Review Letters}\ }\textbf {\bibinfo {volume} {112}},\ \bibinfo
  {pages} {156801} (\bibinfo {year} {2014})},\ \Eprint
  {http://arxiv.org/abs/1309.3571} {arXiv:1309.3571} \BibitemShut {NoStop}%
\bibitem [{\citenamefont {Claassen}\ \emph {et~al.}(2017)\citenamefont
  {Claassen}, \citenamefont {Jiang}, \citenamefont {Moritz},\ and\
  \citenamefont {Devereaux}}]{Claassen2017a}%
  \BibitemOpen
  \bibfield  {author} {\bibinfo {author} {\bibfnamefont {M.}~\bibnamefont
  {Claassen}}, \bibinfo {author} {\bibfnamefont {H.~C.}\ \bibnamefont {Jiang}},
  \bibinfo {author} {\bibfnamefont {B.}~\bibnamefont {Moritz}}, \ and\ \bibinfo
  {author} {\bibfnamefont {T.~P.}\ \bibnamefont {Devereaux}},\ }\href {\doibase
  10.1038/s41467-017-00876-y} {\bibfield  {journal} {\bibinfo  {journal}
  {Nature Communications}\ }\textbf {\bibinfo {volume} {8}},\ \bibinfo {pages}
  {1192} (\bibinfo {year} {2017})},\ \Eprint {http://arxiv.org/abs/1611.07964}
  {arXiv:1611.07964} \BibitemShut {NoStop}%
\bibitem [{\citenamefont {Singla}\ \emph {et~al.}(2015)\citenamefont {Singla},
  \citenamefont {Cotugno}, \citenamefont {Kaiser}, \citenamefont {F{\"{o}}rst},
  \citenamefont {Mitrano}, \citenamefont {Liu}, \citenamefont {Cartella},
  \citenamefont {Manzoni}, \citenamefont {Okamoto}, \citenamefont {Hasegawa},
  \citenamefont {Clark}, \citenamefont {Jaksch},\ and\ \citenamefont
  {Cavalleri}}]{Singla2015}%
  \BibitemOpen
  \bibfield  {author} {\bibinfo {author} {\bibfnamefont {R.}~\bibnamefont
  {Singla}}, \bibinfo {author} {\bibfnamefont {G.}~\bibnamefont {Cotugno}},
  \bibinfo {author} {\bibfnamefont {S.}~\bibnamefont {Kaiser}}, \bibinfo
  {author} {\bibfnamefont {M.}~\bibnamefont {F{\"{o}}rst}}, \bibinfo {author}
  {\bibfnamefont {M.}~\bibnamefont {Mitrano}}, \bibinfo {author} {\bibfnamefont
  {H.~Y.}\ \bibnamefont {Liu}}, \bibinfo {author} {\bibfnamefont
  {A.}~\bibnamefont {Cartella}}, \bibinfo {author} {\bibfnamefont
  {C.}~\bibnamefont {Manzoni}}, \bibinfo {author} {\bibfnamefont
  {H.}~\bibnamefont {Okamoto}}, \bibinfo {author} {\bibfnamefont
  {T.}~\bibnamefont {Hasegawa}}, \bibinfo {author} {\bibfnamefont {S.~R.}\
  \bibnamefont {Clark}}, \bibinfo {author} {\bibfnamefont {D.}~\bibnamefont
  {Jaksch}}, \ and\ \bibinfo {author} {\bibfnamefont {A.}~\bibnamefont
  {Cavalleri}},\ }\href {\doibase 10.1103/PhysRevLett.115.187401} {\bibfield
  {journal} {\bibinfo  {journal} {Physical Review Letters}\ }\textbf {\bibinfo
  {volume} {115}},\ \bibinfo {pages} {187401} (\bibinfo {year} {2015})},\
  \Eprint {http://arxiv.org/abs/1409.1088} {arXiv:1409.1088} \BibitemShut
  {NoStop}%
\bibitem [{\citenamefont {Pomarico}\ \emph {et~al.}(2017)\citenamefont
  {Pomarico}, \citenamefont {Mitrano}, \citenamefont {Bromberger},
  \citenamefont {Sentef}, \citenamefont {Al-Temimy}, \citenamefont {Coletti},
  \citenamefont {St{\"{o}}hr}, \citenamefont {Link}, \citenamefont {Starke},
  \citenamefont {Cacho}, \citenamefont {Chapman}, \citenamefont {Springate},
  \citenamefont {Cavalleri},\ and\ \citenamefont {Gierz}}]{Pomarico2017}%
  \BibitemOpen
  \bibfield  {author} {\bibinfo {author} {\bibfnamefont {E.}~\bibnamefont
  {Pomarico}}, \bibinfo {author} {\bibfnamefont {M.}~\bibnamefont {Mitrano}},
  \bibinfo {author} {\bibfnamefont {H.}~\bibnamefont {Bromberger}}, \bibinfo
  {author} {\bibfnamefont {M.~A.}\ \bibnamefont {Sentef}}, \bibinfo {author}
  {\bibfnamefont {A.}~\bibnamefont {Al-Temimy}}, \bibinfo {author}
  {\bibfnamefont {C.}~\bibnamefont {Coletti}}, \bibinfo {author} {\bibfnamefont
  {A.}~\bibnamefont {St{\"{o}}hr}}, \bibinfo {author} {\bibfnamefont
  {S.}~\bibnamefont {Link}}, \bibinfo {author} {\bibfnamefont {U.}~\bibnamefont
  {Starke}}, \bibinfo {author} {\bibfnamefont {C.}~\bibnamefont {Cacho}},
  \bibinfo {author} {\bibfnamefont {R.}~\bibnamefont {Chapman}}, \bibinfo
  {author} {\bibfnamefont {E.}~\bibnamefont {Springate}}, \bibinfo {author}
  {\bibfnamefont {A.}~\bibnamefont {Cavalleri}}, \ and\ \bibinfo {author}
  {\bibfnamefont {I.}~\bibnamefont {Gierz}},\ }\href {\doibase
  10.1103/PhysRevB.95.024304} {\bibfield  {journal} {\bibinfo  {journal}
  {Physical Review B}\ }\textbf {\bibinfo {volume} {95}},\ \bibinfo {pages}
  {024304} (\bibinfo {year} {2017})},\ \Eprint
  {http://arxiv.org/abs/1607.02314} {arXiv:1607.02314} \BibitemShut {NoStop}%
\bibitem [{\citenamefont {Sentef}(2017)}]{Sentef2017c}%
  \BibitemOpen
  \bibfield  {author} {\bibinfo {author} {\bibfnamefont {M.~A.}\ \bibnamefont
  {Sentef}},\ }\href {\doibase 10.1103/PhysRevB.95.205111} {\bibfield
  {journal} {\bibinfo  {journal} {Physical Review B}\ }\textbf {\bibinfo
  {volume} {95}},\ \bibinfo {pages} {205111} (\bibinfo {year}
  {2017})}\BibitemShut {NoStop}%
\bibitem [{\citenamefont {Coulthard}\ \emph {et~al.}(2017)\citenamefont
  {Coulthard}, \citenamefont {Clark}, \citenamefont {Al-Assam}, \citenamefont
  {Cavalleri},\ and\ \citenamefont {Jaksch}}]{Coulthard2017a}%
  \BibitemOpen
  \bibfield  {author} {\bibinfo {author} {\bibfnamefont {J.~R.}\ \bibnamefont
  {Coulthard}}, \bibinfo {author} {\bibfnamefont {S.~R.}\ \bibnamefont
  {Clark}}, \bibinfo {author} {\bibfnamefont {S.}~\bibnamefont {Al-Assam}},
  \bibinfo {author} {\bibfnamefont {A.}~\bibnamefont {Cavalleri}}, \ and\
  \bibinfo {author} {\bibfnamefont {D.}~\bibnamefont {Jaksch}},\ }\href
  {\doibase 10.1103/PhysRevB.96.085104} {\bibfield  {journal} {\bibinfo
  {journal} {Physical Review B}\ }\textbf {\bibinfo {volume} {96}},\ \bibinfo
  {pages} {085104} (\bibinfo {year} {2017})},\ \Eprint
  {http://arxiv.org/abs/1608.03964} {arXiv:1608.03964} \BibitemShut {NoStop}%
\bibitem [{\citenamefont {Claassen}\ \emph {et~al.}(2016)\citenamefont
  {Claassen}, \citenamefont {Jia}, \citenamefont {Moritz},\ and\ \citenamefont
  {Devereaux}}]{Claassen2016}%
  \BibitemOpen
  \bibfield  {author} {\bibinfo {author} {\bibfnamefont {M.}~\bibnamefont
  {Claassen}}, \bibinfo {author} {\bibfnamefont {C.}~\bibnamefont {Jia}},
  \bibinfo {author} {\bibfnamefont {B.}~\bibnamefont {Moritz}}, \ and\ \bibinfo
  {author} {\bibfnamefont {T.~P.}\ \bibnamefont {Devereaux}},\ }\href {\doibase
  10.1038/ncomms13074} {\bibfield  {journal} {\bibinfo  {journal} {Nature
  Communications}\ }\textbf {\bibinfo {volume} {7}},\ \bibinfo {pages} {13074}
  (\bibinfo {year} {2016})},\ \Eprint {http://arxiv.org/abs/1603.04457}
  {arXiv:1603.04457} \BibitemShut {NoStop}%
\bibitem [{\citenamefont {Kitamura}\ \emph {et~al.}(2017)\citenamefont
  {Kitamura}, \citenamefont {Oka},\ and\ \citenamefont {Aoki}}]{Kitamura2017}%
  \BibitemOpen
  \bibfield  {author} {\bibinfo {author} {\bibfnamefont {S.}~\bibnamefont
  {Kitamura}}, \bibinfo {author} {\bibfnamefont {T.}~\bibnamefont {Oka}}, \
  and\ \bibinfo {author} {\bibfnamefont {H.}~\bibnamefont {Aoki}},\ }\href
  {\doibase 10.1103/PhysRevB.96.014406} {\bibfield  {journal} {\bibinfo
  {journal} {Physical Review B}\ }\textbf {\bibinfo {volume} {96}},\ \bibinfo
  {pages} {014406} (\bibinfo {year} {2017})},\ \Eprint
  {http://arxiv.org/abs/1703.04315} {arXiv:1703.04315} \BibitemShut {NoStop}%
\bibitem [{\citenamefont {Kennes}(2018)}]{Kennes2018c}%
  \BibitemOpen
  \bibfield  {author} {\bibinfo {author} {\bibfnamefont {D.~M.}\ \bibnamefont
  {Kennes}},\ }\href {http://arxiv.org/abs/1801.02866} {\  (\bibinfo {year}
  {2018})},\ \Eprint {http://arxiv.org/abs/1801.02866} {arXiv:1801.02866}
  \BibitemShut {NoStop}%
\bibitem [{\citenamefont {Kennes}\ \emph
  {et~al.}(2018{\natexlab{a}})\citenamefont {Kennes}, \citenamefont {{De La
  Torre}}, \citenamefont {Ron}, \citenamefont {Hsieh},\ and\ \citenamefont
  {Millis}}]{Kennes2018b}%
  \BibitemOpen
  \bibfield  {author} {\bibinfo {author} {\bibfnamefont {D.~M.}\ \bibnamefont
  {Kennes}}, \bibinfo {author} {\bibfnamefont {A.}~\bibnamefont {{De La
  Torre}}}, \bibinfo {author} {\bibfnamefont {A.}~\bibnamefont {Ron}}, \bibinfo
  {author} {\bibfnamefont {D.}~\bibnamefont {Hsieh}}, \ and\ \bibinfo {author}
  {\bibfnamefont {A.~J.}\ \bibnamefont {Millis}},\ }\href {\doibase
  10.1103/PhysRevLett.120.127601} {\bibfield  {journal} {\bibinfo  {journal}
  {Physical Review Letters}\ }\textbf {\bibinfo {volume} {120}},\ \bibinfo
  {pages} {127601} (\bibinfo {year} {2018}{\natexlab{a}})},\ \Eprint
  {http://arxiv.org/abs/1801.06885} {arXiv:1801.06885} \BibitemShut {NoStop}%
\bibitem [{\citenamefont {Tancogne-Dejean}\ \emph {et~al.}(2017)\citenamefont
  {Tancogne-Dejean}, \citenamefont {Sentef},\ and\ \citenamefont
  {Rubio}}]{Tancogne-Dejean2017c}%
  \BibitemOpen
  \bibfield  {author} {\bibinfo {author} {\bibfnamefont {N.}~\bibnamefont
  {Tancogne-Dejean}}, \bibinfo {author} {\bibfnamefont {M.~A.}\ \bibnamefont
  {Sentef}}, \ and\ \bibinfo {author} {\bibfnamefont {A.}~\bibnamefont
  {Rubio}},\ }\href {\doibase 10.1016/B978-0-12-803581-8.09533-3} {\  (\bibinfo
  {year} {2017}),\ 10.1016/B978-0-12-803581-8.09533-3},\ \Eprint
  {http://arxiv.org/abs/1712.01067} {arXiv:1712.01067} \BibitemShut {NoStop}%
\bibitem [{\citenamefont {Schmitt}\ \emph {et~al.}(2008)\citenamefont
  {Schmitt}, \citenamefont {Kirchmann}, \citenamefont {Bovensiepen},
  \citenamefont {Moore}, \citenamefont {Rettig}, \citenamefont {Krenz},
  \citenamefont {Chu}, \citenamefont {Ru}, \citenamefont {Perfetti},
  \citenamefont {Lu}, \citenamefont {Wolf}, \citenamefont {Fisher},\ and\
  \citenamefont {Shen}}]{Schmitt2008}%
  \BibitemOpen
  \bibfield  {author} {\bibinfo {author} {\bibfnamefont {F.}~\bibnamefont
  {Schmitt}}, \bibinfo {author} {\bibfnamefont {P.~S.}\ \bibnamefont
  {Kirchmann}}, \bibinfo {author} {\bibfnamefont {U.}~\bibnamefont
  {Bovensiepen}}, \bibinfo {author} {\bibfnamefont {R.~G.}\ \bibnamefont
  {Moore}}, \bibinfo {author} {\bibfnamefont {L.}~\bibnamefont {Rettig}},
  \bibinfo {author} {\bibfnamefont {M.}~\bibnamefont {Krenz}}, \bibinfo
  {author} {\bibfnamefont {J.-H.}\ \bibnamefont {Chu}}, \bibinfo {author}
  {\bibfnamefont {N.}~\bibnamefont {Ru}}, \bibinfo {author} {\bibfnamefont
  {L.}~\bibnamefont {Perfetti}}, \bibinfo {author} {\bibfnamefont {D.~H.}\
  \bibnamefont {Lu}}, \bibinfo {author} {\bibfnamefont {M.}~\bibnamefont
  {Wolf}}, \bibinfo {author} {\bibfnamefont {I.~R.}\ \bibnamefont {Fisher}}, \
  and\ \bibinfo {author} {\bibfnamefont {Z.-X.}\ \bibnamefont {Shen}},\ }\href
  {http://science.sciencemag.org/content/321/5896/1649.abstract} {\bibfield
  {journal} {\bibinfo  {journal} {Science}\ }\textbf {\bibinfo {volume}
  {321}},\ \bibinfo {pages} {1649 LP } (\bibinfo {year} {2008})}\BibitemShut
  {NoStop}%
\bibitem [{\citenamefont {Matsunaga}\ \emph {et~al.}(2013)\citenamefont
  {Matsunaga}, \citenamefont {Hamada}, \citenamefont {Makise}, \citenamefont
  {Uzawa}, \citenamefont {Terai}, \citenamefont {Wang},\ and\ \citenamefont
  {Shimano}}]{Matsunaga2013}%
  \BibitemOpen
  \bibfield  {author} {\bibinfo {author} {\bibfnamefont {R.}~\bibnamefont
  {Matsunaga}}, \bibinfo {author} {\bibfnamefont {Y.~I.}\ \bibnamefont
  {Hamada}}, \bibinfo {author} {\bibfnamefont {K.}~\bibnamefont {Makise}},
  \bibinfo {author} {\bibfnamefont {Y.}~\bibnamefont {Uzawa}}, \bibinfo
  {author} {\bibfnamefont {H.}~\bibnamefont {Terai}}, \bibinfo {author}
  {\bibfnamefont {Z.}~\bibnamefont {Wang}}, \ and\ \bibinfo {author}
  {\bibfnamefont {R.}~\bibnamefont {Shimano}},\ }\href {\doibase
  10.1103/PhysRevLett.111.057002} {\bibfield  {journal} {\bibinfo  {journal}
  {Physical Review Letters}\ }\textbf {\bibinfo {volume} {111}},\ \bibinfo
  {pages} {057002} (\bibinfo {year} {2013})},\ \Eprint
  {http://arxiv.org/abs/1305.0381} {arXiv:1305.0381} \BibitemShut {NoStop}%
\bibitem [{\citenamefont {Mansart}\ \emph {et~al.}(2013)\citenamefont
  {Mansart}, \citenamefont {Lorenzana}, \citenamefont {Mann}, \citenamefont
  {Odeh}, \citenamefont {Scarongella}, \citenamefont {Chergui},\ and\
  \citenamefont {Carbone}}]{Mansart2013}%
  \BibitemOpen
  \bibfield  {author} {\bibinfo {author} {\bibfnamefont {B.}~\bibnamefont
  {Mansart}}, \bibinfo {author} {\bibfnamefont {J.}~\bibnamefont {Lorenzana}},
  \bibinfo {author} {\bibfnamefont {A.}~\bibnamefont {Mann}}, \bibinfo {author}
  {\bibfnamefont {A.}~\bibnamefont {Odeh}}, \bibinfo {author} {\bibfnamefont
  {M.}~\bibnamefont {Scarongella}}, \bibinfo {author} {\bibfnamefont
  {M.}~\bibnamefont {Chergui}}, \ and\ \bibinfo {author} {\bibfnamefont
  {F.}~\bibnamefont {Carbone}},\ }\href {\doibase 10.1073/pnas.1218742110}
  {\bibfield  {journal} {\bibinfo  {journal} {Proceedings of the National
  Academy of Sciences}\ }\textbf {\bibinfo {volume} {110}},\ \bibinfo {pages}
  {4539} (\bibinfo {year} {2013})},\ \Eprint {http://arxiv.org/abs/1112.0737}
  {arXiv:1112.0737} \BibitemShut {NoStop}%
\bibitem [{\citenamefont {Krauser}\ \emph {et~al.}(2014)\citenamefont
  {Krauser}, \citenamefont {Ebling}, \citenamefont {Fl{\"{a}}schner},
  \citenamefont {Heinze}, \citenamefont {Sengstock}, \citenamefont
  {Lewenstein}, \citenamefont {Eckardt},\ and\ \citenamefont
  {Becker}}]{Krauser2014}%
  \BibitemOpen
  \bibfield  {author} {\bibinfo {author} {\bibfnamefont {J.~S.}\ \bibnamefont
  {Krauser}}, \bibinfo {author} {\bibfnamefont {U.}~\bibnamefont {Ebling}},
  \bibinfo {author} {\bibfnamefont {N.}~\bibnamefont {Fl{\"{a}}schner}},
  \bibinfo {author} {\bibfnamefont {J.}~\bibnamefont {Heinze}}, \bibinfo
  {author} {\bibfnamefont {K.}~\bibnamefont {Sengstock}}, \bibinfo {author}
  {\bibfnamefont {M.}~\bibnamefont {Lewenstein}}, \bibinfo {author}
  {\bibfnamefont {A.}~\bibnamefont {Eckardt}}, \ and\ \bibinfo {author}
  {\bibfnamefont {C.}~\bibnamefont {Becker}},\ }\href {\doibase
  10.1126/science.1244059} {\bibfield  {journal} {\bibinfo  {journal}
  {Science}\ }\textbf {\bibinfo {volume} {343}},\ \bibinfo {pages} {157}
  (\bibinfo {year} {2014})},\ \Eprint {http://arxiv.org/abs/1307.8392}
  {arXiv:1307.8392} \BibitemShut {NoStop}%
\bibitem [{\citenamefont {Murakami}\ \emph {et~al.}(2016)\citenamefont
  {Murakami}, \citenamefont {Werner}, \citenamefont {Tsuji},\ and\
  \citenamefont {Aoki}}]{Murakami2016}%
  \BibitemOpen
  \bibfield  {author} {\bibinfo {author} {\bibfnamefont {Y.}~\bibnamefont
  {Murakami}}, \bibinfo {author} {\bibfnamefont {P.}~\bibnamefont {Werner}},
  \bibinfo {author} {\bibfnamefont {N.}~\bibnamefont {Tsuji}}, \ and\ \bibinfo
  {author} {\bibfnamefont {H.}~\bibnamefont {Aoki}},\ }\href {\doibase
  10.1103/PhysRevB.93.094509} {\bibfield  {journal} {\bibinfo  {journal}
  {Physical Review B}\ }\textbf {\bibinfo {volume} {93}},\ \bibinfo {pages}
  {094509} (\bibinfo {year} {2016})},\ \Eprint
  {http://arxiv.org/abs/1511.06105} {arXiv:1511.06105} \BibitemShut {NoStop}%
\bibitem [{\citenamefont {Sentef}\ \emph {et~al.}(2017)\citenamefont {Sentef},
  \citenamefont {Tokuno}, \citenamefont {Georges},\ and\ \citenamefont
  {Kollath}}]{Sentef2017b}%
  \BibitemOpen
  \bibfield  {author} {\bibinfo {author} {\bibfnamefont {M.~A.}\ \bibnamefont
  {Sentef}}, \bibinfo {author} {\bibfnamefont {A.}~\bibnamefont {Tokuno}},
  \bibinfo {author} {\bibfnamefont {A.}~\bibnamefont {Georges}}, \ and\
  \bibinfo {author} {\bibfnamefont {C.}~\bibnamefont {Kollath}},\ }\href
  {\doibase 10.1103/PhysRevLett.118.087002} {\bibfield  {journal} {\bibinfo
  {journal} {Physical Review Letters}\ }\textbf {\bibinfo {volume} {118}},\
  \bibinfo {pages} {087002} (\bibinfo {year} {2017})},\ \Eprint
  {http://arxiv.org/abs/1611.04307} {arXiv:1611.04307} \BibitemShut {NoStop}%
\bibitem [{\citenamefont {Bukov}\ \emph {et~al.}(2015)\citenamefont {Bukov},
  \citenamefont {D'Alessio},\ and\ \citenamefont {Polkovnikov}}]{Bukov2015c}%
  \BibitemOpen
  \bibfield  {author} {\bibinfo {author} {\bibfnamefont {M.}~\bibnamefont
  {Bukov}}, \bibinfo {author} {\bibfnamefont {L.}~\bibnamefont {D'Alessio}}, \
  and\ \bibinfo {author} {\bibfnamefont {A.}~\bibnamefont {Polkovnikov}},\
  }\href {\doibase 10.1080/00018732.2015.1055918} {\bibfield  {journal}
  {\bibinfo  {journal} {Advances in Physics}\ }\textbf {\bibinfo {volume}
  {64}},\ \bibinfo {pages} {139} (\bibinfo {year} {2015})},\ \Eprint
  {http://arxiv.org/abs/1407.4803} {arXiv:1407.4803} \BibitemShut {NoStop}%
\bibitem [{\citenamefont {Abanin}\ \emph {et~al.}(2017)\citenamefont {Abanin},
  \citenamefont {{De Roeck}}, \citenamefont {Ho},\ and\ \citenamefont
  {Huveneers}}]{Abanin2017}%
  \BibitemOpen
  \bibfield  {author} {\bibinfo {author} {\bibfnamefont {D.~A.}\ \bibnamefont
  {Abanin}}, \bibinfo {author} {\bibfnamefont {W.}~\bibnamefont {{De Roeck}}},
  \bibinfo {author} {\bibfnamefont {W.~W.}\ \bibnamefont {Ho}}, \ and\ \bibinfo
  {author} {\bibfnamefont {F.}~\bibnamefont {Huveneers}},\ }\href {\doibase
  10.1103/PhysRevB.95.014112} {\bibfield  {journal} {\bibinfo  {journal}
  {Physical Review B}\ }\textbf {\bibinfo {volume} {95}},\ \bibinfo {pages}
  {014112} (\bibinfo {year} {2017})},\ \Eprint
  {http://arxiv.org/abs/1510.03405} {arXiv:1510.03405} \BibitemShut {NoStop}%
\bibitem [{\citenamefont {Emery}(1987)}]{Emery1987}%
  \BibitemOpen
  \bibfield  {author} {\bibinfo {author} {\bibfnamefont {V.~J.}\ \bibnamefont
  {Emery}},\ }\href {\doibase 10.1103/PhysRevLett.58.2794} {\bibfield
  {journal} {\bibinfo  {journal} {Physical Review Letters}\ }\textbf {\bibinfo
  {volume} {58}},\ \bibinfo {pages} {2794} (\bibinfo {year}
  {1987})}\BibitemShut {NoStop}%
\bibitem [{\citenamefont {Keimer}\ \emph {et~al.}(2015)\citenamefont {Keimer},
  \citenamefont {Kivelson}, \citenamefont {Norman}, \citenamefont {Uchida},\
  and\ \citenamefont {Zaanen}}]{Keimer2015}%
  \BibitemOpen
  \bibfield  {author} {\bibinfo {author} {\bibfnamefont {B.}~\bibnamefont
  {Keimer}}, \bibinfo {author} {\bibfnamefont {S.~A.}\ \bibnamefont
  {Kivelson}}, \bibinfo {author} {\bibfnamefont {M.~R.}\ \bibnamefont
  {Norman}}, \bibinfo {author} {\bibfnamefont {S.}~\bibnamefont {Uchida}}, \
  and\ \bibinfo {author} {\bibfnamefont {J.}~\bibnamefont {Zaanen}},\ }\href
  {\doibase 10.1038/nature14165} {\enquote {\bibinfo {title} {{From quantum
  matter to high-temperature superconductivity in copper oxides}},}\ }
  (\bibinfo {year} {2015})\BibitemShut {NoStop}%
\bibitem [{\citenamefont {Mentink}\ \emph {et~al.}(2015)\citenamefont
  {Mentink}, \citenamefont {Balzer},\ and\ \citenamefont
  {Eckstein}}]{Mentink2015a}%
  \BibitemOpen
  \bibfield  {author} {\bibinfo {author} {\bibfnamefont {J.~H.}\ \bibnamefont
  {Mentink}}, \bibinfo {author} {\bibfnamefont {K.}~\bibnamefont {Balzer}}, \
  and\ \bibinfo {author} {\bibfnamefont {M.}~\bibnamefont {Eckstein}},\ }\href
  {\doibase 10.1038/ncomms7708} {\bibfield  {journal} {\bibinfo  {journal}
  {Nature Communications}\ }\textbf {\bibinfo {volume} {6}},\ \bibinfo {pages}
  {6708} (\bibinfo {year} {2015})},\ \Eprint
  {http://arxiv.org/abs/arXiv:1407.4761v1} {arXiv:arXiv:1407.4761v1}
  \BibitemShut {NoStop}%
\bibitem [{\citenamefont {Peronaci}\ \emph {et~al.}(2018)\citenamefont
  {Peronaci}, \citenamefont {Schir{\'{o}}},\ and\ \citenamefont
  {Parcollet}}]{Peronaci2018}%
  \BibitemOpen
  \bibfield  {author} {\bibinfo {author} {\bibfnamefont {F.}~\bibnamefont
  {Peronaci}}, \bibinfo {author} {\bibfnamefont {M.}~\bibnamefont
  {Schir{\'{o}}}}, \ and\ \bibinfo {author} {\bibfnamefont {O.}~\bibnamefont
  {Parcollet}},\ }\href {\doibase 10.1103/PhysRevLett.120.197601} {\bibfield
  {journal} {\bibinfo  {journal} {Physical Review Letters}\ }\textbf {\bibinfo
  {volume} {120}},\ \bibinfo {pages} {197601} (\bibinfo {year} {2018})},\
  \Eprint {http://arxiv.org/abs/1711.07889} {arXiv:1711.07889} \BibitemShut
  {NoStop}%
\bibitem [{\citenamefont {Magnus}(1954)}]{Magnus1954a}%
  \BibitemOpen
  \bibfield  {author} {\bibinfo {author} {\bibfnamefont {W.}~\bibnamefont
  {Magnus}},\ }\href {\doibase 10.1002/cpa.3160070404} {\bibfield  {journal}
  {\bibinfo  {journal} {Communications on Pure and Applied Mathematics}\
  }\textbf {\bibinfo {volume} {7}},\ \bibinfo {pages} {649} (\bibinfo {year}
  {1954})}\BibitemShut {NoStop}%
\bibitem [{\citenamefont {Bucksbaum}\ \emph {et~al.}(1990)\citenamefont
  {Bucksbaum}, \citenamefont {Zavriyev}, \citenamefont {Muller},\ and\
  \citenamefont {Schumacher}}]{Bucksbaum1990}%
  \BibitemOpen
  \bibfield  {author} {\bibinfo {author} {\bibfnamefont {P.~H.}\ \bibnamefont
  {Bucksbaum}}, \bibinfo {author} {\bibfnamefont {A.}~\bibnamefont {Zavriyev}},
  \bibinfo {author} {\bibfnamefont {H.~G.}\ \bibnamefont {Muller}}, \ and\
  \bibinfo {author} {\bibfnamefont {D.~W.}\ \bibnamefont {Schumacher}},\ }\href
  {\doibase 10.1103/PhysRevLett.64.1883} {\bibfield  {journal} {\bibinfo
  {journal} {Physical Review Letters}\ }\textbf {\bibinfo {volume} {64}},\
  \bibinfo {pages} {1883} (\bibinfo {year} {1990})}\BibitemShut {NoStop}%
\bibitem [{\citenamefont {Eckardt}(2017)}]{Eckardt2017a}%
  \BibitemOpen
  \bibfield  {author} {\bibinfo {author} {\bibfnamefont {A.}~\bibnamefont
  {Eckardt}},\ }\href {\doibase 10.1103/RevModPhys.89.011004} {\bibfield
  {journal} {\bibinfo  {journal} {Reviews of Modern Physics}\ }\textbf
  {\bibinfo {volume} {89}},\ \bibinfo {pages} {011004} (\bibinfo {year}
  {2017})},\ \Eprint {http://arxiv.org/abs/1606.08041} {arXiv:1606.08041}
  \BibitemShut {NoStop}%
\bibitem [{\citenamefont {Ohta}\ \emph {et~al.}(1991)\citenamefont {Ohta},
  \citenamefont {Tohyama},\ and\ \citenamefont {Maekawa}}]{Ohta1991}%
  \BibitemOpen
  \bibfield  {author} {\bibinfo {author} {\bibfnamefont {Y.}~\bibnamefont
  {Ohta}}, \bibinfo {author} {\bibfnamefont {T.}~\bibnamefont {Tohyama}}, \
  and\ \bibinfo {author} {\bibfnamefont {S.}~\bibnamefont {Maekawa}},\ }\href
  {\doibase 10.1103/PhysRevB.43.2968} {\bibfield  {journal} {\bibinfo
  {journal} {Physical Review B}\ }\textbf {\bibinfo {volume} {43}},\ \bibinfo
  {pages} {2968} (\bibinfo {year} {1991})}\BibitemShut {NoStop}%
\bibitem [{\citenamefont {Chen}\ \emph {et~al.}(2010)\citenamefont {Chen},
  \citenamefont {Moritz}, \citenamefont {Vernay}, \citenamefont {Hancock},
  \citenamefont {Johnston}, \citenamefont {Jia}, \citenamefont
  {Chabot-Couture}, \citenamefont {Greven}, \citenamefont {Elfimov},
  \citenamefont {Sawatzky},\ and\ \citenamefont {Devereaux}}]{Chen2010c}%
  \BibitemOpen
  \bibfield  {author} {\bibinfo {author} {\bibfnamefont {C.~C.}\ \bibnamefont
  {Chen}}, \bibinfo {author} {\bibfnamefont {B.}~\bibnamefont {Moritz}},
  \bibinfo {author} {\bibfnamefont {F.}~\bibnamefont {Vernay}}, \bibinfo
  {author} {\bibfnamefont {J.~N.}\ \bibnamefont {Hancock}}, \bibinfo {author}
  {\bibfnamefont {S.}~\bibnamefont {Johnston}}, \bibinfo {author}
  {\bibfnamefont {C.~J.}\ \bibnamefont {Jia}}, \bibinfo {author} {\bibfnamefont
  {G.}~\bibnamefont {Chabot-Couture}}, \bibinfo {author} {\bibfnamefont
  {M.}~\bibnamefont {Greven}}, \bibinfo {author} {\bibfnamefont
  {I.}~\bibnamefont {Elfimov}}, \bibinfo {author} {\bibfnamefont {G.~A.}\
  \bibnamefont {Sawatzky}}, \ and\ \bibinfo {author} {\bibfnamefont {T.~P.}\
  \bibnamefont {Devereaux}},\ }\href {\doibase 10.1103/PhysRevLett.105.177401}
  {\bibfield  {journal} {\bibinfo  {journal} {Physical Review Letters}\
  }\textbf {\bibinfo {volume} {105}},\ \bibinfo {pages} {177401} (\bibinfo
  {year} {2010})},\ \Eprint {http://arxiv.org/abs/1002.4683} {arXiv:1002.4683}
  \BibitemShut {NoStop}%
\bibitem [{\citenamefont {Chen}\ \emph {et~al.}(2013)\citenamefont {Chen},
  \citenamefont {Sentef}, \citenamefont {Kung}, \citenamefont {Jia},
  \citenamefont {Thomale}, \citenamefont {Moritz}, \citenamefont {Kampf},\ and\
  \citenamefont {Devereaux}}]{Chen2013}%
  \BibitemOpen
  \bibfield  {author} {\bibinfo {author} {\bibfnamefont {C.~C.}\ \bibnamefont
  {Chen}}, \bibinfo {author} {\bibfnamefont {M.}~\bibnamefont {Sentef}},
  \bibinfo {author} {\bibfnamefont {Y.~F.}\ \bibnamefont {Kung}}, \bibinfo
  {author} {\bibfnamefont {C.~J.}\ \bibnamefont {Jia}}, \bibinfo {author}
  {\bibfnamefont {R.}~\bibnamefont {Thomale}}, \bibinfo {author} {\bibfnamefont
  {B.}~\bibnamefont {Moritz}}, \bibinfo {author} {\bibfnamefont {A.~P.}\
  \bibnamefont {Kampf}}, \ and\ \bibinfo {author} {\bibfnamefont {T.~P.}\
  \bibnamefont {Devereaux}},\ }\href {\doibase 10.1103/PhysRevB.87.165144}
  {\bibfield  {journal} {\bibinfo  {journal} {Physical Review B - Condensed
  Matter and Materials Physics}\ }\textbf {\bibinfo {volume} {87}},\ \bibinfo
  {pages} {165144} (\bibinfo {year} {2013})}\BibitemShut {NoStop}%
\bibitem [{\citenamefont {Metzner}\ \emph {et~al.}(2012)\citenamefont
  {Metzner}, \citenamefont {Salmhofer}, \citenamefont {Honerkamp},
  \citenamefont {Meden},\ and\ \citenamefont
  {Sch{\"{o}}nhammer}}]{Metzner2012a}%
  \BibitemOpen
  \bibfield  {author} {\bibinfo {author} {\bibfnamefont {W.}~\bibnamefont
  {Metzner}}, \bibinfo {author} {\bibfnamefont {M.}~\bibnamefont {Salmhofer}},
  \bibinfo {author} {\bibfnamefont {C.}~\bibnamefont {Honerkamp}}, \bibinfo
  {author} {\bibfnamefont {V.}~\bibnamefont {Meden}}, \ and\ \bibinfo {author}
  {\bibfnamefont {K.}~\bibnamefont {Sch{\"{o}}nhammer}},\ }\href {\doibase
  10.1103/RevModPhys.84.299} {\bibfield  {journal} {\bibinfo  {journal}
  {Reviews of Modern Physics}\ }\textbf {\bibinfo {volume} {84}},\ \bibinfo
  {pages} {299} (\bibinfo {year} {2012})},\ \Eprint
  {http://arxiv.org/abs/1105.5289} {arXiv:1105.5289} \BibitemShut {NoStop}%
\bibitem [{\citenamefont {Kennes}\ \emph
  {et~al.}(2018{\natexlab{b}})\citenamefont {Kennes}, \citenamefont
  {Lischner},\ and\ \citenamefont {Karrasch}}]{Kennes2018d}%
  \BibitemOpen
  \bibfield  {author} {\bibinfo {author} {\bibfnamefont {D.~M.}\ \bibnamefont
  {Kennes}}, \bibinfo {author} {\bibfnamefont {J.}~\bibnamefont {Lischner}}, \
  and\ \bibinfo {author} {\bibfnamefont {C.}~\bibnamefont {Karrasch}},\ }\href
  {http://arxiv.org/abs/1805.06310} {\  (\bibinfo {year}
  {2018}{\natexlab{b}})},\ \Eprint {http://arxiv.org/abs/1805.06310}
  {arXiv:1805.06310} \BibitemShut {NoStop}%
\bibitem [{\citenamefont {Maeno}\ \emph {et~al.}(1994)\citenamefont {Maeno},
  \citenamefont {Hashimoto}, \citenamefont {Yoshida}, \citenamefont
  {Nishizaki}, \citenamefont {Fujita}, \citenamefont {Bednorz},\ and\
  \citenamefont {Lichtenberg}}]{Maeno1994}%
  \BibitemOpen
  \bibfield  {author} {\bibinfo {author} {\bibfnamefont {Y.}~\bibnamefont
  {Maeno}}, \bibinfo {author} {\bibfnamefont {H.}~\bibnamefont {Hashimoto}},
  \bibinfo {author} {\bibfnamefont {K.}~\bibnamefont {Yoshida}}, \bibinfo
  {author} {\bibfnamefont {S.}~\bibnamefont {Nishizaki}}, \bibinfo {author}
  {\bibfnamefont {T.}~\bibnamefont {Fujita}}, \bibinfo {author} {\bibfnamefont
  {J.~G.}\ \bibnamefont {Bednorz}}, \ and\ \bibinfo {author} {\bibfnamefont
  {F.}~\bibnamefont {Lichtenberg}},\ }\href {\doibase 10.1038/372532a0}
  {\bibfield  {journal} {\bibinfo  {journal} {Nature}\ }\textbf {\bibinfo
  {volume} {372}},\ \bibinfo {pages} {532} (\bibinfo {year} {1994})},\ \Eprint
  {http://arxiv.org/abs/0611691} {arXiv:0611691 [cond-mat]} \BibitemShut
  {NoStop}%
\bibitem [{\citenamefont {Cao}\ \emph {et~al.}(2018)\citenamefont {Cao},
  \citenamefont {Fatemi}, \citenamefont {Fang}, \citenamefont {Watanabe},
  \citenamefont {Taniguchi}, \citenamefont {Kaxiras},\ and\ \citenamefont
  {Jarillo-Herrero}}]{Cao2018c}%
  \BibitemOpen
  \bibfield  {author} {\bibinfo {author} {\bibfnamefont {Y.}~\bibnamefont
  {Cao}}, \bibinfo {author} {\bibfnamefont {V.}~\bibnamefont {Fatemi}},
  \bibinfo {author} {\bibfnamefont {S.}~\bibnamefont {Fang}}, \bibinfo {author}
  {\bibfnamefont {K.}~\bibnamefont {Watanabe}}, \bibinfo {author}
  {\bibfnamefont {T.}~\bibnamefont {Taniguchi}}, \bibinfo {author}
  {\bibfnamefont {E.}~\bibnamefont {Kaxiras}}, \ and\ \bibinfo {author}
  {\bibfnamefont {P.}~\bibnamefont {Jarillo-Herrero}},\ }\href {\doibase
  10.1038/nature26160} {\bibfield  {journal} {\bibinfo  {journal} {Nature}\
  }\textbf {\bibinfo {volume} {556}},\ \bibinfo {pages} {43} (\bibinfo {year}
  {2018})},\ \Eprint {http://arxiv.org/abs/1803.02342} {arXiv:1803.02342}
  \BibitemShut {NoStop}%
\end{thebibliography}%

\clearpage\newpage
\begin{widetext}
\begin{center}\large{\bf Supplementary Information:\\
 Light-induced d-wave superconductivity through
Floquet-engineered Fermi surfaces in cuprates}
\end{center}
\begin{figure}[b]
	\centering
	\includegraphics[width=6cm]{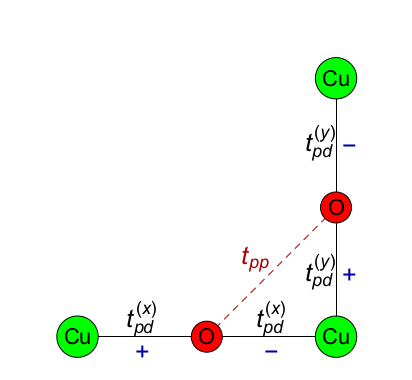}
	\caption{Schematic depiction of the anisotropic frustrated Cu-O cluster that serves as a starting point for a strong-coupling expansion. Signs $\pm$ illustrate the sign structure of Cu-O hopping matrix elements.}
	\label{fig:CuOcluster}
\end{figure}

\section{Strong-Coupling Perturbation Theory for Cuprate Three-Orbital Models}

Here, we describe the strong-coupling perturbation theory for a frustrated anisotropic three-band model of the Cu $d_{x^2-y^2}$ orbital and O $p_x$, $p_y$ orbitals. In the main text, anisotropy arises from unidirectional motional narrowing due to optical pumping in the high-frequency limit -- we therefore consider an effective Floquet three-band model with static (as opposed to time-dependent) anistropy.

Since the strong-coupling expansion generates purely local spin-exchange couplings, it is sufficient to start from a minimal five-site cluster of three Cu atoms and two O sites, to arrive at a frustrated $J, J'$ Heisenberg model at half filling. Fig. \ref{fig:CuOcluster} depicts the corresponding cluster geometry, with Hamiltonian
\begin{align}
	\Ham = \Ham_{\rm O} + \Ham_{\rm Cu-O} + \Ham_{\rm int}
\end{align}
with
\begin{align}
	\Ham_{\rm O} &= \sum_{\substack{n={\rm H,V} \\ \sigma}} \Delta~ \hat{p}_{n\sigma}^\dag \hat{p}_{n\sigma} - t_{pp} \sum_{\sigma} \left( \hat{p}^\dag_{{\rm H}\sigma} \hat{p}_{{\rm V}\sigma} + \hc \right) \\
	\Ham_{\rm Cu-O} &= - \left[ t^{(x)}_{pd} \left( \hat{d}^\dag_{1\sigma} - \hat{d}^\dag_{2\sigma} \right) \hat{p}_{{\rm H}\sigma} + t^{(y)}_{pd} \left( \left( \hat{d}^\dag_{2\sigma} - \hat{d}^\dag_{3\sigma} \right) \hat{p}_{{\rm V}\sigma} \right) +  \hc \right] \\
	\Ham_{\rm int} &= U_{dd} \sum_{\substack{i=1,2,3 \\ \sigma}} \hat{d}^\dag_{i\uparrow} \hat{d}_{i\uparrow} \hat{d}^\dag_{i\downarrow} \hat{d}_{i\downarrow} + U_{pp} \sum_{\substack{n={\rm H,V} \\ \sigma}} \hat{p}^\dag_{n\uparrow} \hat{p}_{n\uparrow} \hat{p}^\dag_{n\downarrow} \hat{p}_{n\downarrow}
\end{align}
Here, $\Ham_{\rm O}$, $\Ham_{\rm Cu-O}$ and $\Ham_{\rm int}$ describe kinetic contributions from O sites, Cu-O hybridization, and interactions, respectively. $\Delta$ is the charge transfer energy for O sites, $t_{pd}$ and $t_{pp}$ are Cu-O and O-O hopping matrix elements, respectively, and $U_{dd}$, $U_{pp}$ are local Coulomb repulsion matrix elements on Cu and O sites.

To fourth order in a straightforward perturbative expansion in powers of $t_{pd}^{(x,y)}$, one obtains an anisotropic nearest-neighbor Heisenberg spin exchange model between local moments on Cu without frustration. Conversely, any dependence on $t_{pp}$ would be incorporated only to higher orders in the perturbation series.

To circumvent the necessity for higher-order perturbative expansions, we therefore first diagonalize the kinetic coupling between O sites, subsequently perform a fourth-order strong-coupling expansion \textit{in the rotated basis}. Define the hybridized O orbitals of Fig. \ref{fig:CuOcluster} as
\begin{align}
	\hat{a}_{\pm,\sigma} = \frac{1}{\sqrt{2}} \left( \hat{p}_{{\rm H} \sigma} \pm \hat{p}_{{\rm V} \sigma} \right)
\end{align}
This choice diagonalizes $\Ham_{\rm O}$ while transforming the Cu-O hybridization matrix elements such that $t_{pp}$ enters on equal footing as the charge transfer energy and interaction energy scales. Now, a fourth-order expansion can be performed straightforwardly in powers of the transformed $H_{\rm Cu-O}$, where the Cu orbitals are taken to be singly-occupied whereas the O sites are filled/empty in electron/hole language. Tedious but straightforward perturbation theory finally recovers the $J-J'$ model discussed in the main text (yielding reasonable results for the hopping elements in equilibrium as a sanity check), which is used to derive a corresponding anisotropic single-band Hubbard model, as an effective starting point for FRG. 

\end{widetext}
\end{document}